\documentclass[prd,onecolumn,nofootinbib,superscriptaddress,showkeys,citeautoscrip]{revtex4}
\pdfoutput=1
\usepackage{graphicx}
\usepackage{epstopdf}
\usepackage{amsmath}
\usepackage{hyperref}
\usepackage{cleveref}
\crefname{equation}{Eq.}{Eqs.}
\crefname{figure}{Fig.}{Figs.}
\crefname{table}{Table}{Tables}
\newcommand{\crefrangeconjunction}{--}
\usepackage{dcolumn}
\usepackage{fancybox}
\usepackage{bm}
\usepackage{multirow}
\usepackage[usenames,dvipsnames]{color}
\usepackage{array}
\usepackage{rotating}
\usepackage{ulem,fancyvrb}
\usepackage{latexsym}
\usepackage{amsfonts}
\usepackage{xcolor}
\usepackage{longtable}
\usepackage{amssymb}
\usepackage{xspace}
\def\tana2{\tan^2\alpha_H}
\def\tanb2{\tan^2\beta}

\def\m0pr{m_0'}

\def\Mz2{M_Z^2}

\def\GeV{~{\rm GeV}}
\def\TeV{~{\rm TeV}}
\def\met100{\slashed{E}_T\geq 100~{\rm GeV}}

\newcommand{\beqn}{\begin{eqnarray}}
\newcommand{\eeqn}{\end{eqnarray}}
\newcommand{\be}{\begin{equation}}
\newcommand{\ee}{\end{equation}}
\def \cha{\tilde{\chi}^{\pm}_1}

\def \n34{\tilde{\chi}^{0}_{3,4}}
\newcommand{\g}{\ensuremath{\tilde{g}}}

\def \ta{\tilde{t}_1}

\def \ba{\tilde{b}_1}

\def \mhf{m_{1/2}}
\newcommand{\Ft}[1]{\ensuremath{\tilde{F}_{#1}}}
\newcommand{\Ht}[1]{\ensuremath{\tilde{H}_{#1}}}
\newcommand{\Gt}[1]{\ensuremath{\tilde{G}_{#1}}}
\newcommand{\mH}[1]{\ensuremath{m_{H_{#1}}^2}}
\newcommand{\et}{\ensuremath{\tilde{e}}}
\newcommand{\ft}{\ensuremath{\tilde{f}}}
\newcommand{\Zf}[1]{\ensuremath{Z_{#1}^f}}
\newcommand{\Zh}[1]{\ensuremath{Z_{#1}^h}}
\newcommand{\Dt}{\ensuremath{D(t)}}
\newcommand{\MHt}[1]{\ensuremath{M_{\tilde{H}_{#1}}}}
\newcommand{\MFt}[1]{\ensuremath{M_{\tilde{F}_{#1}}}}
\newcommand{\MGt}[1]{\ensuremath{M_{\tilde{G}_{#1}}}}

\newcommand{\Met}{\ensuremath{M_{\tilde{e}}}}
\newcommand{\Mft}{\ensuremath{M_{\tilde{f}}}}
\newcommand{\vMHt}[1]{\ensuremath{\vec{M}_{\tilde{H}_{#1}}}}

\newcommand{\vMft}{\ensuremath{\vec{M}_{\tilde{f}}}}
\newcommand{\MmH}[1]{\ensuremath{M_{m_{H_{#1}}}}}

\def \atG{{\tilde{\alpha}}_G}
\def \dint{\displaystyle\int_0^t}
\def \dpsty{\displaystyle}
\def \tbeta{\tan\beta}

\newcommand{\sta}{\ensuremath{\tilde{\tau}_1}}

\def \smr{\tilde{\mu}_R}
\def \ser{\tilde{e}_R~}

\def \mhu {m^2_{H_u}}

\def \pd {p \rightarrow \overline{\nu} K^+}

\begin{document}

\title{Higgs Boson Mass,   Proton Decay,   Naturalness
and   Constraints of LHC and Planck Data}

\author{Mengxi~Liu\footnote{Email: m.liu@neu.edu }}
\affiliation{Department of Physics, Northeastern University,
 Boston, MA 02115, USA}

\author{Pran~Nath\footnote{Email: nath@neu.edu}}
\affiliation{Department of Physics, Northeastern University,
 Boston, MA 02115, USA}

\date{\today}

 \begin{abstract}
 A Higgs boson mass $\sim 126$ GeV as determined by the LHC data requires a large loop correction
 which in turn implies a large sfermion mass.  Implication of this result for the stability of the proton 
 in supersymmetric  grand unified theories is examined including other experiments constraints 
 along with the most recent result on cold dark matter from Planck.  It is shown that over the allowed
 parameter space of supergravity unified models, proton lifetime is highly sensitive to the
 Higgs boson mass and a few GeV shift in its mass can change the proton decay lifetime for the
 mode $p\to \bar \nu K^+$ by as much as two orders of magnitude or more. An analysis is also
 given on the nature of radiative breaking of the electroweak symmetry 
 in view of the high Higgs boson, and it is shown that
 most of the parameter space of universal and non-universal supergravity unified models  lies on 
 the Hyperbolic Branch of radiative breaking of the electroweak symmetry, while the 
 Ellipsoidal Branch and the Focal Point regions are highly depleted and contain only a 
 very small region of the allowed parameter space.  Also discussed are the naturalness 
 criteria when the proton stability constraints along with the electroweak symmetry breaking 
 are considered  together. It is shown that under the assumed naturalness criteria the overall fine tuning is 
 improved for larger values of the scalar mass with the inclusion of the proton stability constraint. Thus
the naturalness criteria including proton stability along with 
electroweak symmetry breaking constraints tend to favor the weak scale of SUSY in the several TeV region.
 Implications for the discovery of supersymmetry in view of the high Higgs mass 
 are briefly discussed. 

\end{abstract}

\maketitle

\section{Introduction \label{intro}}
 Over the past year the ATLAS and the CMS Collaborations have identified a signal for a boson 
 around $\sim 126$ GeV. Thus the  ATLAS Collaboration finds a signal at 
$126.0 \pm 0.4 ({\rm stat})\pm 0.4({\rm sys})~{\rm GeV}$
which is  at the $5.0\sigma$ level~\cite{:2012gk} while the  CMS Collaboration finds a  signal at 
$125.3\pm 0.4 ({\rm stat})\pm 0.5({\rm sys})~{\rm GeV}$
at the $5.0\sigma$ level~\cite{:2012gu}.
While the properties of the new boson still need to be fully established, it is widely believed that
the discovered boson is indeed the Higgs boson~\cite{Englert:1964et,Higgs:1964pj,Guralnik:1964eu}
 that enters in the breaking of the electroweak symmetry of the Standard Model ~\cite{Weinberg:1967tq,salam}.
 Remarkably the Higgs boson mass lies close to the upper limit predicted in  supergravity grand 
 unified models~\cite{Chamseddine:1982jx,Nath:1983aw,Hall:1983iz,Arnowitt:1992aq}
 which predict an upper limit of around $130$ GeV~\cite{Akula:2011aa,Akula:2012kk,Arbey:2012dq,Ellis:2012aa,Baer:2012mv}
(For a recent review of Higgs and supersymmetry see~\cite{Nath:2012nh}).
The  high mass  $\sim 126$ GeV requires a large loop correction which in turn implies  that some 
 of the sparticles entering the loop corrections (for a review see~\cite{Carena:2002es}) 
  to the Higgs mass must be in the several TeV range.  In this case
 the heavy particles could be out of reach of the LHC. One possibility is that a part of the Higgs boson 
 arises from sources outside of the MSSM such as from corrections arising from vector like 
 multiplets\cite{Babu:2008ge,Martin:2010dc,Feng:2013mea,Joglekar:2013zya}. 
 However, in this work we do not make that assumption.  \\

 In the early analyses  using radiative breaking of the electroweak symmetry(for a review see~\cite{Ibanez:2007pf}) 
 only the Ellipsoidal Branch was known, in that a fixed value of the $\mu$ (the Higgs mixing parameter) implied
 upper limits on sparticle masses. However, the situation changed drastically with the discovery of the 
 Hyperbolic Branch ~\cite{Chan:1997bi,Chattopadhyay:2003xi} 
 (for related work see~\cite{bbbkt,Feldman:2011ud}) 
 when it was discovered that another branch of radiative breaking
 of the electroweak symmetry existed where the sparticle masses could  lie in the several TeV region
 while $\mu$ could still be at the sub TeV scale. Specifically on this branch TeV size scalars can exist consistent with 
 small $\mu$. 
 In this work we investigate  the allowed parameter space of
 supergravity models under the constraint that the models accommodate the high Higgs mass.
 We show that for supergravity models  most of the allowed parameter
 space under the high Higgs mass restriction lies on the Hyperbolic Branch while the Ellipsoidal 
 Branch and Focal Point region accommodate only  a small fraction of the allowed parameter space. 
We discuss the above for supergravity models with universal boundary conditions (mSUGRA/CMSSM)
as well as 
supergravity models with non-universal gaugino masses (NuSUGRA).  Sensitivity of the proton lifetime to the 
Higgs boson mass is investigated and it is shown that the proton lifetime is correlated very sensitively
to the Higgs boson mass. Further we discuss  issues of naturalness in view of the large Higgs boson
mass and the stability of the proton. 
It is shown that a composite fine tuning 
including  proton stability along with the radiative electroweak symmetry breaking constraint  prefers
a  SUSY scale in the several TeV region.  \\

The outline of the rest of the paper is as follows:
 In Sec.(\ref{SecEWSB}) we discuss the radiative breaking of the electroweak symmetry under the constraint
 of the high Higgs boson mass. Here we show that most of the parameter space of supergravity unified 
 models with universal boundary conditions lies on the Hyperbolic branch while the Ellipsoidal Branch
 and the Focal Point region are essentially empty.   In Sec.(\ref{Secproton}) we discuss the implications 
 of the high Higgs boson mass on the proton lifetime and show that the proton lifetime is very sensitive
 to small shifts in the Higgs boson mass. Thus a shift of a  few GeV of the light Higgs boson mass 
 can change the proton lifetime by as muc\emph{•}h as two orders of magnitude or more. 
 In Sec.(\ref{NonUni}) we extend the discussion to supergravity unified models with non-universalities 
 and show that the broad  conclusions drawn in the previous sections still hold. 
 In Sec.(\ref{ft}) we discuss the issue of naturalness and fine tuning  when the proton stability constraints
 are combined with the constraints from electroweak symmetry breaking. Here it is shown that 
 the fine tuning criteria including both the proton stability and the electroweak symmetry breaking constraints
 favor a high sfermion scale.  Conclusions are given in Sec.(\ref{SecConclu}).

 \section{Higgs mass and  branches of radiative breaking of the electroweak symmetry\label{SecEWSB}}
 It is of interest to investigate the allowed parameter space of the supergravity unified models under the constraint
 of the high Higgs boson mass.  We consider first supergravity  unified models with universal boundary 
 conditions consisting of the universal scalar mass $m_0$, universal gaugino mass $m_{1/2}$, 
 universal trilinear coupling $A_0$, $\tan\beta =<H_2>/<H_1>$ where $H_2$ gives mass to the 
 up quarks and $H_1$ gives mass to the down quarks and leptons, and the Higgs mixing parameter
 $\mu$ which enters the superpotential via the term $\mu H_1H_2$.
 Of specific interest is to determine the  branch of radiative breaking of the 
electroweak symmetry preferred by the  high mass. 
 Thus the radiative electroweak
 symmetry breaking can be exhibited in the following form~\cite{Nath:1997qm,Chan:1997bi}

\beqn
 \mu^2 +\frac{1}{2}M_Z^2 =   m^2_0  C_1+ A^{'2}_0 C_2  + m^2_{1/2} C_3'+ \Delta \mu^2_{\rm loop}~,
\label{1.1}
\eeqn
where  $A_o'\equiv A_0 + \frac{C_4}{2C_2} m_{1/2}$ and

\beqn \label{1.2}
C_1=\frac{1}{\tan^2\beta-1}\left(1-\frac{3 D_0-1}{2}\tan^2\beta\right)~,
 C_2=\frac{\tan^2\beta}{\tan^2\beta-1}k~,
C_3'\equiv  C_3 -\frac{C_4^2}{4C_2}, \\ 
C_3=\frac{1}{\tan^2\beta-1}\left(g- e\tan^2\beta \right)~, 
C_4=-\frac{\tan^2\beta}{\tan^2\beta-1}f~.
\label{1.3}
\eeqn
 Here $e.f,g,k$ are as defined in~\cite{Ibanez:1984vq} and $D_0(t)$ is  defined by 
\beqn
 D_0(t)= \left(1+ 6 Y_0 F(t)\right)^{-1}~.
 \label{1.4}
 \eeqn
In the above $Y_0 =h_t(0)^2/(4\pi^2)$, where $h_t(0)$ is the top Yukawa coupling at the GUT scale, $M_G\simeq2\times10^{16}\GeV$. 
$F(t)$ is defined by  $F(t) = \int_0^t E(t^\prime) dt^\prime~,$ where  
$E(t)=\left(1 + \beta_3 t\right)^{16/3b_3} \left(1+ \beta_2 t\right)^{3/b_2} \left(1+ \beta_1t\right)^{13/9 b_1}$.
Here 
$\beta_i = \alpha_i(0) b_i/(4\pi)$ and $b_i=(-3, 1, 11)$ for $SU(3), SU(2)$ and $U(1)$
 and $t= \ln \left(M_G^2/Q^2\right)$ where $Q$ is the
 renormalization group point.  We are using the normalizations where  $\alpha_3(0) = \alpha_2(0) = \frac{5}{3} \alpha_1(0)
 =\alpha_G(0)$ and $\alpha_G(0)$ is the common value of the normalized $\alpha's$ at the GUT scale. 
 Finally, $\Delta \mu^2_{\rm loop}$ is the loop correction~\cite{Arnowitt:1992qp}.  To understand the 
 origin of the branches of radiative breaking it is useful to choose a renormalization group scale
 $Q$ where the loop correction $\Delta \mu^2_{\rm loop}$ is minimized. 
 In this circumstance if all the coefficients $C_1, C_2, C_3'$ are positive, the right hand side of Eq.(\ref{1.1}) is a
positive  sum of squares which leads to an upper limit on each of soft parameters determined by the size of 
$\mu^2 + \frac{1}{2} M_Z^2$ on the left hand side. This is the so called Ellipsoidal Branch (EB) where 
$\mu$  sets an upper limit on the soft parameters and thus on the size of the sparticle masses.
This is typically the case if the loop correction $\Delta \mu^2_{\rm loop}$ is small. However, the situation changes 
drastically if the loop correction $\Delta \mu^2_{\rm loop}$ is large. This is so because $C_i$  are  
functions of the  
renormalization group (RG) scale Q  and for the case when the loop correction $\Delta \mu^2_{\rm loop}$ is large
the RG dependence of $C_i$ can become significant. Indeed as we change the renormalization group
scale $Q$, there is a rapid change in  $\Delta \mu^2_{\rm loop}$,   and a rapid 
 compensating change  also in the remaining terms on the right hand side of Eq.(\ref{1.1}) so that 
$\mu^2$ does not exhibit any rapid dependence on $Q$. Now it turns out that there are regions of the
parameter space where one or more of the $C_i$ may turn negative as $Q$  varies. For the supergravity unified
models with universal boundary conditions this is the case for $C_1$, i.e., in certain regions of the parameter
space $C_1$ can turn negative while the remainder on the right hand side of Eq.(\ref{1.1}) remains positive.
In this case it is useful to write Eq.(\ref{1.1}) in the following form 
  \beqn
 \mu^2  = 
 \begin{pmatrix}
\dpsty +1	~~~~({\rm EB})\\
  ~0	~~~~({\rm FP})\\
  -1 ~~~~({\rm HB})\\
\end{pmatrix}
   m^2_0  |C_1|+  \Delta^2 ,
\label{1.5}
\eeqn
where $\Delta^2$ stands for the rest of the terms in Eq.(\ref{1.1}).
In Eq.(\ref{1.5})  $+1$ corresponds to the Ellipsoidal Branch (EB), $-1$ corresponds to the Hyperbolic Branch (HB) and 
$C_1=0$ is the boundary point between the two which we call Focal Point (FP). Its approximate form 
when $\tan\beta >>1$ is the Focus Point~\cite{Feng:1999mn}. $C_1=0$ is achieved when 
$D_0 = 1/3$  (see Appendix A ).
We wish now to identify the allowed regions of the mSUGRA parameter space in terms of the branch on which they
reside, i.e., EB, HB or FP.
 To quantify the region FP we define a small corridor  around  $C_1=0$.
This is feasible since FP is very sensitive to the top quark mass and we utilize the error in the top quark 
mass to define the corridor around $C_1=0$. Currently the top quark mass is determined to be 
$m_t = (173.5\pm 1.0)\GeV$ and thus we define the FP corridor so that  
 \cite{Akula:2011jx}, 
\be
|C_1| <   \delta\left(Q,m_{t}\right), ~~~\delta\left(Q,m_{t}\right)\ll1~,
\label{1.6}
\ee
where 
\be
\delta (Q, m_t)  \simeq 3\left(1-D_0\right) \frac{\delta m_t}{m_t}~,
\label{1.7}
\ee
and where $D_0$  is defined in Eq.(\ref{1.4}).
Thus the Focal Point corresponds to the corridor  $-\vert \delta \vert < C_1 < \vert \delta\vert$, 
the EB  corresponds to $C_1 > \vert \delta \vert$  and HB corresponds to $C_1 < -\vert \delta \vert$.
EB consists of closed elliptical curves and closed surfaces in the soft parameter space for fixed $\mu$, 
while the HB region  $C_1 <  - \vert \delta \vert$ consists of open curves and open surfaces. 
We now define a focal curve (FC) on HB as the one where two soft parameters can get large while
$\mu$ remains fixed. It was shown in~\cite{Akula:2011jx}
that in mSUGRA there exist two varieties of 
Focal Curves FC1 and FC2 as shown in Table I. On FC1, $m_{1/2}$ and $\mu$  remain fixed
while $m_0$ and $m_{1/2}$ get large, and thus FC1 is an open curve lying in the $m_0-A_0$ plane. 
On FC2, $A_0$ and $\mu$  remain fixed
while $m_0$ and $A_0$ get large, and thus FC2 is an open curve lying in the $m_0-m_{1/2}$ plane. 
A convolution of focal curves leads to focal surfaces~\footnote{
The classification of the parameter space of SUGRA models into focal curves and focal surfaces
is a geometric one independent of issues of fine tuning. The focal curves and focal surfaces  automatically
arise on HB for the mSUGRA case when $C_1<0$.  For NuSUGRA, the HB 
gets redefined such that $\mu$ remains constant while 
two or more soft parameters get large due to one or more of the $C_i$  turning negative as discussed
in Sec. IV.}.
It is interesting to classify the allowed parameter space  of mSUGRA in terms of the branch of radiative breaking 
of the electroweak symmetry
they
lie on, i.e., EB, HB or FP.  This is done 
under the constraints of the most recent LHC searches~\cite{cmsREACH,AtlasSUSY,atlas0lep,atlas165pb,atlas1fb} and other experimental constraints including the most recent results from the Planck experiment~\cite{planck}.

\begin{table}[h!]
\begin{center}
\begin{tabular}{|c|c|c|}
\hline
 Focal Curve   & Large soft parameters  & Small soft parameters \\
\hline
HB/FC1 & $m_0-A_0$  & $m_{1/2}$ \\
HB/FC2 & $m_0-m_{1/2}$  & $A_0$ \\
  \hline
\end{tabular}
\label{tab_msugra}
\caption{Classification of focal curves in mSUGRA.
The focal curve HB/FC1 corresponds to the case when 
$m_{1/2}$ is kept fixed while $m_0$ and $A_0$ get large
keeping $\mu$ fixed (The asymptotic form of these curves give $m_0/A_0=\pm 1$\cite{Feldman:2011ud}).
The focal curve HB/FC2 corresponds to the case
when $A_0$ and $\mu$ are kept fixed while $m_0$ and $m_{1/2}$ get large.\label{tab_msugra}}
\end{center}
\end{table}

\begin{figure}[t!]
\begin{center}
\includegraphics[scale=0.4]{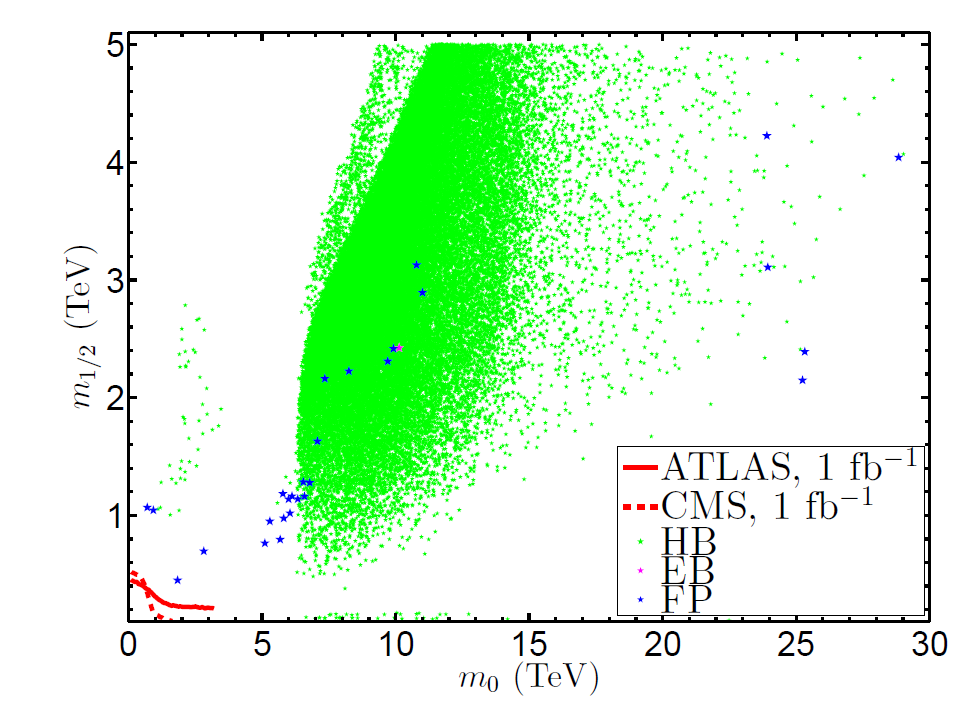}
\caption{\label{mSUGRAlhc}
The parameter points in the $m_0-m_{1/2}$ plane in supergravity unified models with universal
boundary conditions passing the general constraints. The plot exhibits the parameter points 
that lie on HB (green), on EB (red) and on FP (blue). The analysis shows that most of the allowed
parameter space lies on HB while the allowed regions of  EB and FP are essentially empty except for a few scattered 
points. For the analysis here and elsewhere in the paper we have used a top mass of 172.9 GeV. 
The region excluded by the ATLAS and CMS collaborations is also exhibited.}
\end{center}
\end{figure}

We investigate the issue of classification of the branches of mSUGRA 
by mapping the soft parameters space  in the following ranges: $m_0 \in (200\GeV, 30\TeV)$, 
$m_{1/2} \in (100\GeV, 5\TeV)$, 
$A_0 \in (-6m_0,6m_0)$, 
and $\tan\beta \in(1,60)$.
Experimental constraints are then applied for all model points including the limits on sparticle masses from LEP~\cite{pdgrev}:
$m_{\sta} > 81.9\GeV$, 
$m_{\cha} > 103.5\GeV$, 
$m_{\ta} > 95.7\GeV$, 
$m_{\ba} > 89\GeV$, 
$m_{\ser} > 107\GeV$, 
$m_{\smr} > 94\GeV$, 
and $m_{\g} > 308\GeV$.
The most recent Planck measurement~\cite{planck} of the relic density of cold dark matter gives  $\Omega_\chi h^2= 0.1199\pm 0.0027$.
Here we apply the $4\sigma$ upper bound, i.e. $\Omega_\chi h^2 < 0.13$.  
Other constraints applied   include the $g_\mu -2$ 
constraint $\left(-11.4\times 10^{-10}\right)\leq \delta \left(g_{\mu}-2\right) \leq \left(9.4\times10^{-9}\right)$ and  the FCNC constraint from
B-physics~{measurements}~\cite{bphys,cmslhcbbsmumu,Abazov:2010fs}, i.e. 
$\left(2.77\times 10^{-4} \right)\leq {\mathcal{B}r}\left(b\to s\gamma\right) \leq \left( 4.37\times 10^{-4}\right)$  
and
${\mathcal{B}r}\left(B_{s}\to \mu^{+}\mu^-\right)\leq 1.1\times10^{-8}$.  As done in~\cite{Akula:2011ke,Akula:2011jx}, we will refer to these constraints as the  {\it general constraints}.  These constraints are
 imposed using {\sc micrOMEGAs}~\cite{belanger} for the  relic density as well as  for the indirect constraints 
and {\sc SoftSUSY}~\cite{Allanach}   
for the sparticle mass {spectrum}. 
We will also consider NuSUGRA models
(for recent works on NuSUGRA see~\cite{Martin:2009ad,Feldman:2009zc,Gogoladze:2012yf}
and for a review see~\cite{Nath:2010zj}. String based models also allow for non-universalities of gaugino masses,
see, e.g.,~\cite{Kaufman:2013pya}).
The supergravity grand unification formalism of \cite{Chamseddine:1982jx} still applies. 
For the NuSUGRA case to be discussed in Sec.IV all of the experimental constraints discussed
above still apply except that
the  ranges of the soft parameters are chosen as follows:  $m_0 \in (200\GeV, 30\TeV)$, 
$m_{i} \in (100\GeV, 5\TeV)$, 
$A_0 \in (-6m_0,6m_0)$, 
$\tan\beta \in(1,60)$ 
where $i=1,2,3$ for NuSUGRA.
\\

In Fig.\ref{mSUGRAlhc} we exhibit the allowed parameter space of the supergravity unified models
with universal boundary conditions in the $m_0-m_{1/2}$ plane consistent with all the constraints
discussed above. The region excluded by the most recent ATLAS and CMS searches is exhibited.
In this figure we also show the regions of the parameter space that lie on the HB, EB, and FP
branches of radiative breaking of the  electroweak symmetry.  The figure shows that essentially
all the parameter space of the universal supergravity unified model lies in the HB region (indicated by
green points) and the EB  region (indicated by red points) and the FP  region (indicated by blue points) are essentially all empty except for a few scattered points (see also~\cite{Akula:2011jx}).
 
 \section{Proton stability\label{Secproton}}
In supersymmetric GUTs proton decay from dimension five operators 
depends very sensitively 
 on the sparticle spectrum since the  sparticle spectrum enters in the dressing 
 loop diagrams which  involve the exchange of squarks and sleptons, gluinos,  charginos, 
 and  neutralinos\cite{drw,enr,Arnowitt:1985iy,Hisano:1992jj,Goto:1998qg}  
  (for recent reviews see~\cite{Nath:2006ut,Raby:2008pd,Hewett:2012ns}). 
 Thus low values of sfermion masses can lead to too rapid a proton decay 
 for the mode $p\to \bar \nu K^+$   in conflict with the current experimental limit~\cite{Hewett:2012ns}, i.e,
 \beqn
 \tau^{exp}(p\to \bar \nu K^+) > 4 \times 10^{33} yr.
 \eeqn 
Since  a heavy Higgs boson mass in the vicinity of $\sim 126$ GeV implies relatively
large values of sfermion masses 
it is pertinent to investigate  proton stability within the constraint of the experimentally
observed large Higgs boson mass. We will limit ourselves to generic $SU(5)$ type models.
Further, while chargino  $\tilde\chi^{\pm}$, gluino $\tilde g$ and neutralino $\tilde \chi^0$  exchange diagrams all contribute to
the decay width, the dominant contribution comes from the chargino 
exchange diagram
and we will limit ourselves to considerations for decay with this exchange. Thus here
the decay width is given by~\cite{Nath:1997jc}, 
\be
\Gamma(p\rightarrow \bar\nu_iK^+)=(\frac {\beta_p}{M_{H_3}})^2|A|^2
|B_i|^2C ,
\label{3.1}
\ee
where $M_{H_3}$ is the  Higgsino triplet mass and $\beta_p$ is the matrix 
element between the proton and the vacuum state of the 3 quark operator so that
$\beta_p U_L^{\gamma}=\epsilon_{abc}\epsilon_{\alpha \beta} <0|d_{aL}^{\alpha}
u_{bL}^{\beta}u_{cL}^{\gamma}|p>$
where $U^{\gamma}_L$ is the proton spinor.
The most reliable evaluation of $\beta_p$ comes from lattice gauge 
calculations and is given~\cite{Aoki:2006ib} as $\beta_p=0.0118~\text{GeV}^3$.
Other factors that appear in Eq.(\ref{3.1}) have the following meaning: $A$
contains the quark mass and CKM factors, $B_i$  are the functions
that describe the dressing loop diagrams, and $C$ contains chiral Lagrangian
factors which convert  the Lagrangian involving quark fields to the
effective Lagrangian  involving mesons and baryons.
Individually these functions are given by   
\begin{equation}
A=\frac{\alpha_2^2}{2M_W^2}m_s m_c V_{21}^{\dagger} V_{21}A_L A_S,
\end{equation}
where $m_s (m_c)$ are the strange (charm) quark mass,  $V_{ij}$ are the CKM factors, and $A_L$ and $A_S$ are the long 
distance and the short distance 
 renormalization group suppression factors as one  evolves the
operators from the GUT scale down to the electro-weak 
scale and then from the electroweak scale down to 1GeV
~\cite{enr,Dermisek:2000hr,EmmanuelCosta:2003pu,Nihei:1994tx,Turzynski:2001zs}, and 
$B_i$ are given by 
\begin{equation}
B_i= \frac{1}{\sin2\beta}\frac{m_i^d V_{i1}^{\dagger}}{m_sV_{21}^{\dagger}} 
[P_2 B_{2i}+\frac{m_tV_{31}V_{32}}{m_cV_{21}V_{22}} P_3B_{3i}],
\end{equation}
where $m_i^d$ is the down quark mass for flavor $i$ and $m_t$ is the top quark mass.
Here the first term in the bracket is the contribution from the second
generation and the second term is the contribution from the third generation
and $P_2, P_3$ with values ($\pm 1$)  are the relative parities of the second and the 
third generation contributions. 
The functions $B_{ji}$ are the loop integrals defined by
$B_{ji}=F(\tilde u_i,\tilde d_j,\tilde{\chi}^{\pm})+(\tilde d_j\rightarrow \tilde e_j)$, where 

\begin{figure}[t!]
\begin{center}
\includegraphics[scale=0.3]{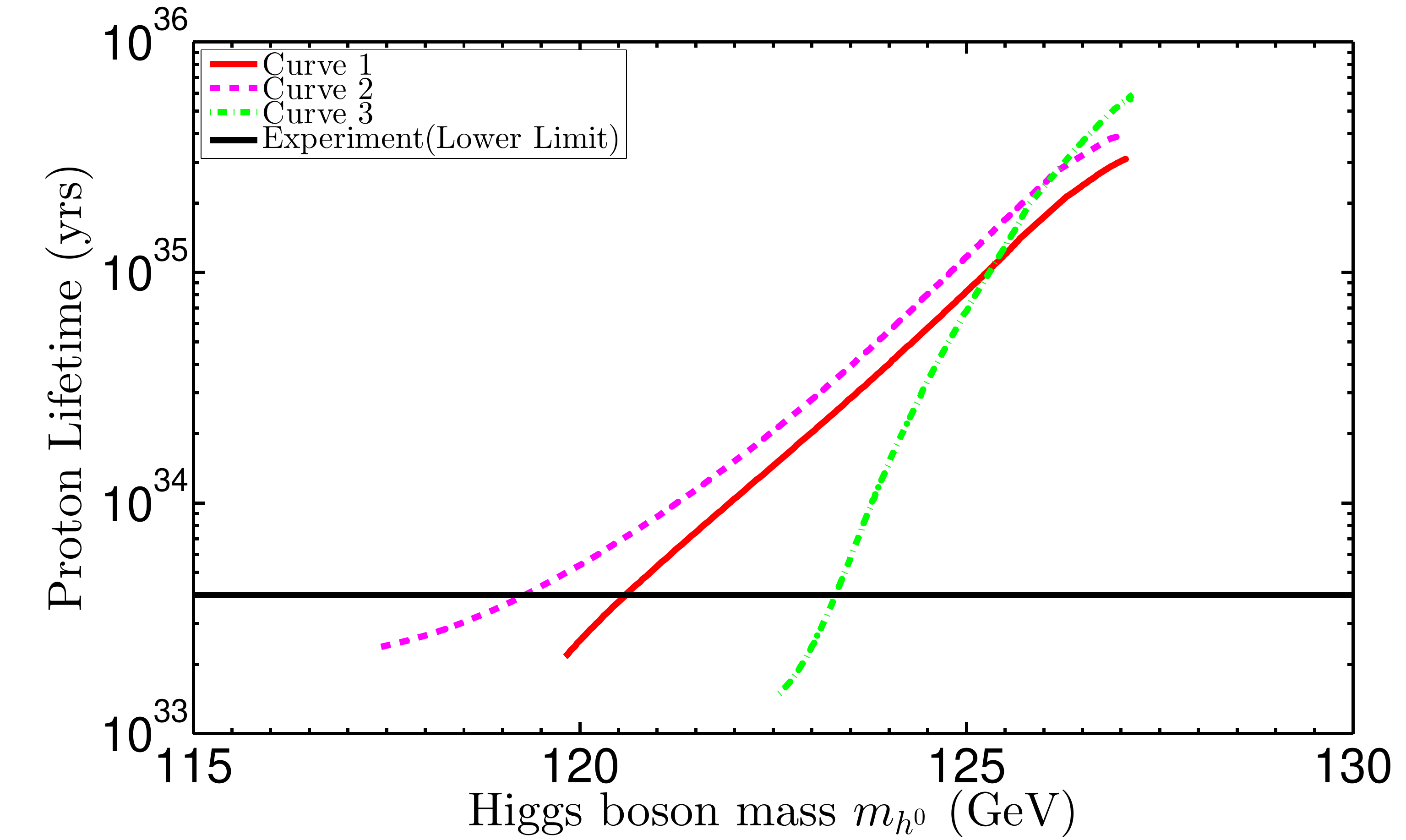}
\caption{\label{ltmhiggs}
An exhibition of the  sensitive dependence of the proton lifetime for the decay mode $p\to \bar \nu K^+$ 
as a function of the Higgs boson mass for the  supergravity unified model  with universal boundary conditions. 
Parameters for curves 1-3 are as follows: Curve 1: $\mhf=   4207\GeV, A_0=  20823\GeV, \tan\beta=   7.3$ while $m_0$ varies, 
$M_{H_3}^{eff}/M_G=50$ here and for other curves;
Curve 2: $\mhf=   2035\GeV, A_0=  16336\GeV, \tan\beta=   8$ while $m_0$ and $A_0$ vary
 and Curve 3: $\mhf=   3048\GeV, A_0 / m_0  =  -0.5, \tan\beta=  6.5$ while $m_0$ and $A_0$ vary.}
\end{center}
\end{figure}

\begin{eqnarray}
F(\tilde u_i,\tilde d_j,\tilde{\chi}^{\pm})=[E \cos\gamma_-\sin\gamma_+\tilde f(\tilde
u_i,\tilde d_j, \tilde{\chi}^{\pm}_1)
+\cos\gamma_+\sin\gamma_-\tilde f(\tilde
u_i,\tilde d_j, \tilde{\chi}^{\pm}_2)]\nonumber\\
-\frac{1}{2} \frac{\delta_{i3}m_i^u \sin2\delta_{ui}}{\sqrt 2 M_W \sin\beta}
[E \sin\gamma_-\sin\gamma_+\tilde f(\tilde
u_{i1},\tilde d_j, \tilde{\chi}^{\pm}_1)
-\cos\gamma_-\cos\gamma_+\tilde f(\tilde
u_{i1},\tilde d_j, \tilde{\chi}^{\pm}_2)\nonumber\\
 - (\tilde u_{i1}\rightarrow \tilde u_{i2})],
 \label{3.5}
\end{eqnarray}
and where $\tilde f$ appearing in Eq.(\ref{3.5}) is given by 
\begin{equation}
\tilde f(\tilde u_i,\tilde d_j, \tilde{\chi}^{\pm}_k)=\sin^2\delta_{ui}
 f(\tilde u_{i1},\tilde d_j, \tilde{\chi}^{\pm}_k)
+\cos^2\delta_{ui}
 f(\tilde u_{i2},\tilde d_j, \tilde{\chi}^{\pm}_k).
\end{equation}
Here  the tilde quantities in the arguments are the sparticle masses, i.e., 
$\tilde u_i$ are the up squark masses for flavor $i$ and $\tilde d_j$ are the down squark masses for flavor $j$ 
and the function $f$ is defined by
\begin{equation}
f(a,b,c)=\frac{m_c}{m_b^2-m_c^2}[\frac{m_b^2}{m_a^2-m_b^2}\ln(\frac{m_a^2}
{m_b^2})-(m_a\rightarrow m_c)].
\end{equation}
Further in Eq.(\ref{3.5}) 
$ \gamma_{\pm}=\beta_+\pm\beta_- $ where 
$ \sin2\beta_{\pm}={(\mu\pm m_2)}/{[4\nu_{\pm}^2
+(\mu\pm m_2)^2]^{1/2}}$, 
$\sqrt 2 \nu_{\pm}=M_W(\sin\beta\pm \cos\beta)$ and 
$\sin2\delta_{u3}=-{2(A_t+\mu \cot\beta)m_t}/({m_{\tilde t_1}^2-
m_{\tilde t_2}^2})$, 
$E=1$ when  $\sin2\beta>\mu m_2/M_W^2$ and
$E=-1$ when $\sin2\beta<\mu m_2/M_W^2$.  
Finally $C$ is given by 
\begin{equation}
C=\frac{m_N}{32\pi f_{\pi}^2} [(1+\frac {m_N(D+F)}{m_B})
(1-\frac{m_K^2}{m_N^2})]^2,
\end{equation}
where 
$\tilde t_i$ are the stop masses and
$f_{\pi}, D,F, ..$ etc are the chiral Lagrangian factors and we use the numerical values
$f_{\pi}=0.131$~GeV,
$D=0.8$,
$F=0.47$,
$m_N$=0.94~GeV,
$m_K$=0.495~GeV, 
$m_B$=1.15~GeV and we choose $P_2=1$ and $P_3=-1$. 
The partial decay lifetime of the proton into $p\to \bar \nu K^+$ mode is given by 
$\tau(\pd)=  {\hbar}/{\Gamma(\pd)}$.   \\

Typically supersymmetric models give too rapid a  proton decay for the mode $p\to \bar \nu K^+$ from
dimension five operators~\cite{Murayama:2001ur}. 
One possible way out 
is the cancellation mechanism for the reduction of proton decay arising from 
different Higgs triplet
representations at the GUT scale~\cite{Nath:2007eg}. 
 This is equivalent to raising the value of the effective Higgs triplet mass. \cite{Arnowitt:1993pd}.
  Specification of the GUT physics allows one to determine the effective Higgs triplet 
  mass (see, e.g.,\cite{Nath:2007eg,Babu:2010ej}).
 Here, however, we do not  commit to a specific GUT structure but rather consider  $SU(5)$ like models 
 where due to various Higgs representations
that enter at the GUT scale one has a number of Higgs triplets/anti-triplets $H_i, \bar H_i$.
Suppose we choose the basis in which only $H_1, \bar H_1$ couple to matter, i.e., one has
couplings of the 
type~\cite{Arnowitt:1993pd} 
    $\bar H_1 J + \bar K H_1 + \bar H_i M_{ij} H_j$, 
where $J$ and $\bar K$ are bilinear in matter fields and $M_{ij}$ is the superheavy Higgs mass matrix. 
Many grand unified models automatically lead to such a possibility~\cite{Babu:2011tw,bgns}.
Specifically in models of the type discussed in ~\cite{Babu:2011tw} one has only one light doublet
and several Higgs triplets/anti-triplets.
On eliminating the superheavy fields one finds that the effective proton decay operator is of the 
form $-\bar K (M_{H_3}')^{-1}J$ where $M_{H_3}' = (M^{-1}_{11})^{-1}$. This allows $M_{H_3}'$ to be much
larger than the GUT scale. In the analysis here we will use the effective mass $M_{H_3}^{eff}=M_{H_3}'/A_LA_S$ and 
we  consider three cases $M_{H_3}^{eff}/M_G=10, 25, 50$ for analysis in this work.\\

\begin{figure}[t!]
\begin{center}
\includegraphics[scale=0.18]{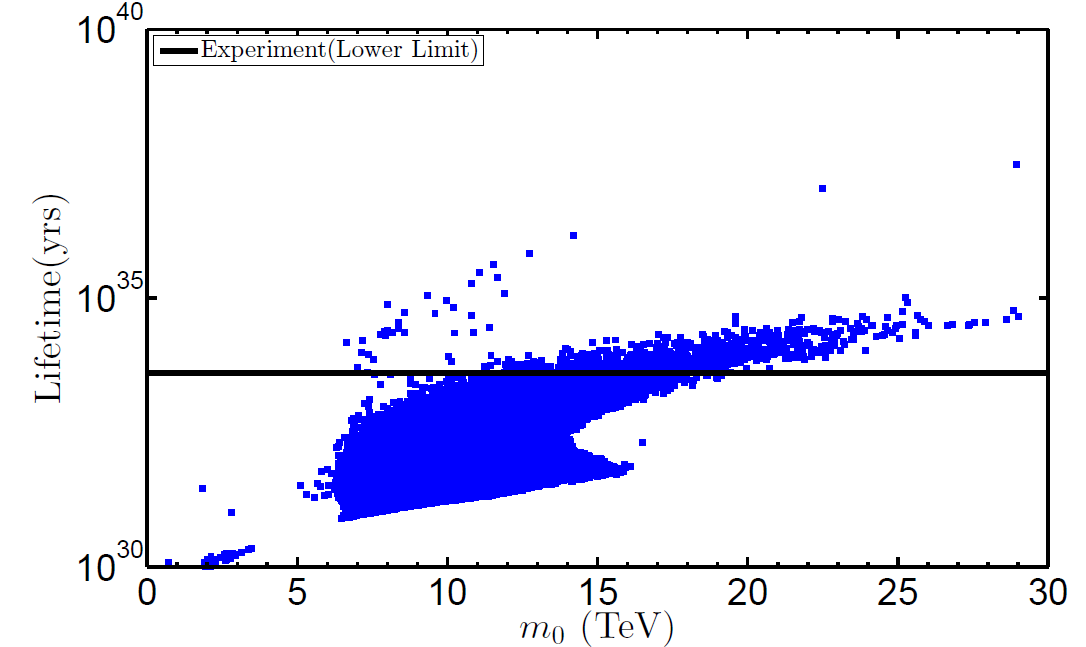}
\includegraphics[scale=0.18]{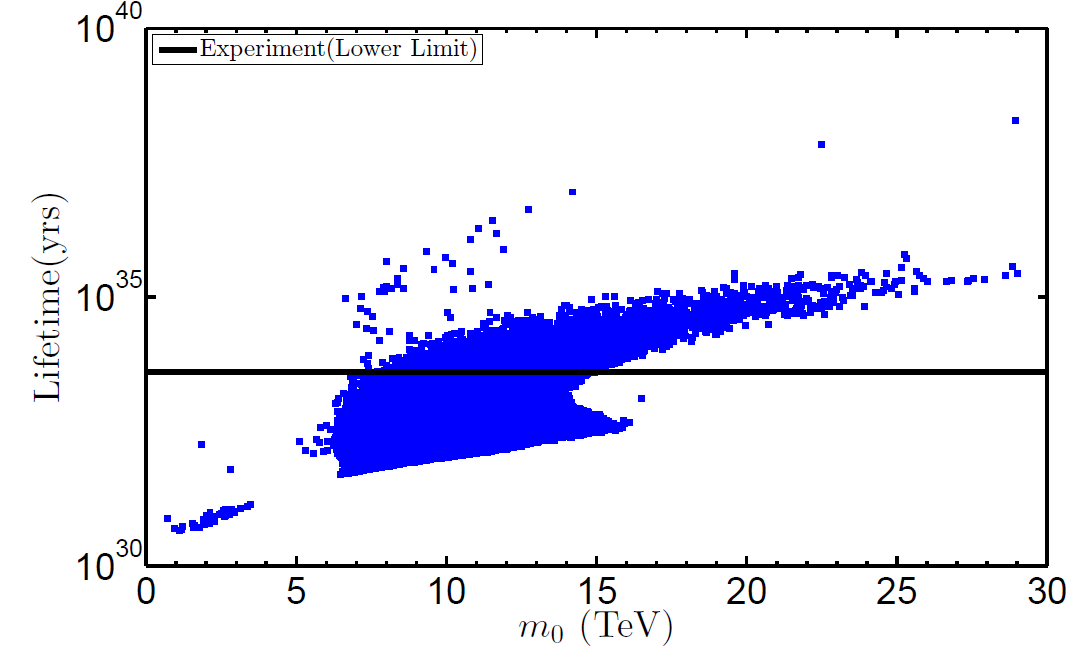}
\includegraphics[scale=0.18]{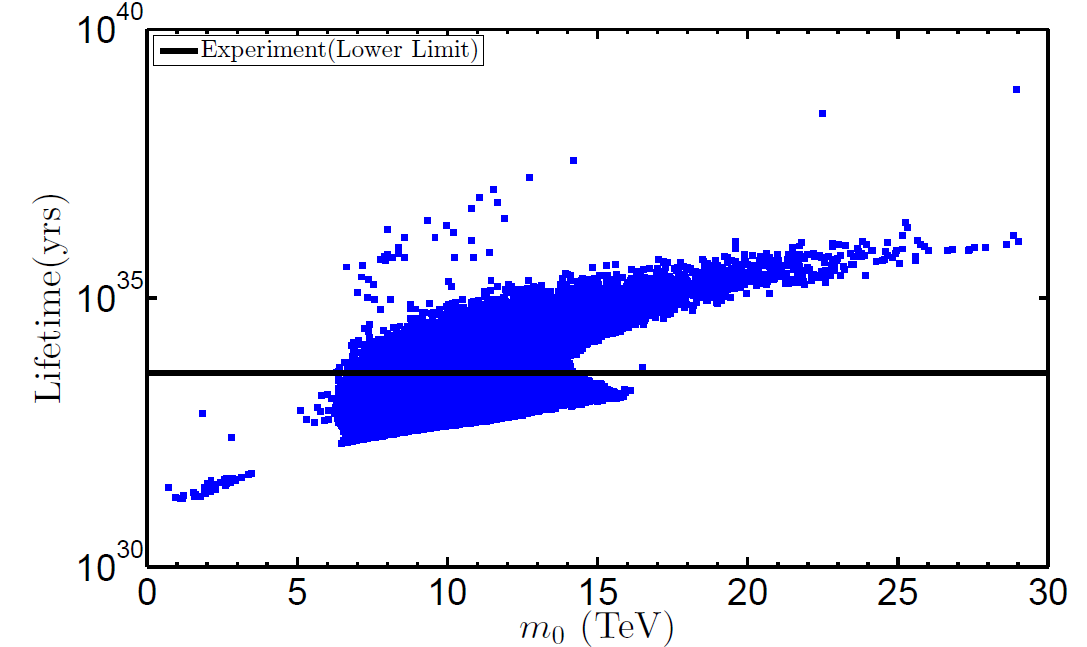}
\caption{\label{ltmsugra}
An exhibition of the partial  lifetime for the decay mode $\pd$  given by blue squares as a function of $m_0$
over the parameter space of the supergravity model  with universal boundary conditions over the allowed ranges  consistent with all the 
experimental constraints. Left panel: The  case when $M_{H_3}^{eff}/M_G=10$. Middle panel: Same as the left panel except for the case $M_{H_3}^{eff}/M_G=25$. 
Right panel: Same as the left panel except for the case $M_{H_3}^{eff}/M_G=50$. The  current experimental lower limit for
 this mode is given by the    horizontal black line. 
 The analysis given here is consistent with the Higgs boson mass within a   $2\sigma$  range.
 }
\end{center}
\end{figure}

In Fig.(\ref{ltmhiggs}) we exhibit the dependence of the proton lifetime for the decay mode $\pd$
as a function of 
the Higgs boson mass under the  constraints discussed in the caption of Fig.(\ref{ltmhiggs}).
The curves show a very
sharp dependence of  the proton lifetime on the Higgs boson mass   
 which increases by up to two
orders of magnitude with a shift in the mass of the Higgs boson in the range of 5-10 GeV. 
In Fig.(\ref{ltmsugra}) we exhibit the proton 
lifetime for the decay mode $\pd$ 
as a function of $m_0$ for the three values of $M_{H_3}^{eff}$
when all the parameters in the model are allowed to vary consistent with the 
radiative electroweak  symmetry breaking constraints and the experimental constraints including
those from the LHC and the Planck experiment. 
  One finds that the parameters compatible with  all the constraints clearly prefer values of $m_0$ in the several TeV region.

\section{NuSUGRA: Focal curves and Focal surfaces\label{NonUni}}

In Sec.(II) a classification of radiative breaking of the electroweak symmetry is given in terms of the branches on which the allowed 
parameter space of mSUGRA resides.  Here we extend the analysis to NuSUGRA and classify the allowed
   parameter space under the constraints of radiative breaking of the electroweak symmetry and  
    all the experimental constraints, i.e.,  
 we discuss the composition of the parameter space in terms of
HB, EP and FP. We will also discuss the sensitivity of the proton decay lifetime for  the mode $p\to \bar\nu K^+$
for the NuSUGRA case. 
For specificity we  define the gaugino masses at the grand unification scale by $m_i$ where
$m_i = m_{1/2} (1+ \delta_i)$, i=1,2,3
and where $\delta_i$ define the non-universalities in the $U(1), SU(2)_L, SU(3)_C$ sectors.
It is shown in Appendix~\ref{coeff_nug} that in this case the radiative electroweak symmetry breaking equation 
Eq.(\ref{1.1}) for the universal soft breaking case is replaced by 

\beqn
\mu^2+\frac{1}{2}M_Z^2 = C_1m_0^2+C_2A_0^2+ \tilde C_3^{ij}m_im_j+\tilde C_4^im_iA_0+\Delta{\mu}^2,
\label{4.1}
\eeqn
where $C_1$ and $C_2$ are as defined by Eq.(\ref{1.2}) while $\tilde C_3^{ij}$ and $\tilde C_4^i$
are given by 
\begin{gather}
\tilde  C_3^{ij}=\frac{\left(\MmH{1}\right)_{ij}-\tan^2\beta\left(\Met\right)_{ij}}{\tan^2\beta-1},
~~\tilde C_4^i= -\frac{\tan^2\beta}{\tan^2\beta-1}\left(M_{\tilde{f}}\right)_i
\label{c4ng}
\end{gather}
Here  $\MmH{1}$, $\Met$ and $\Mft$ are defined in Appendix B. 
$\tilde C_3$ and $\tilde C_4$ in  Eq.(\ref{c4ng}) reduce to the universal case when
$m_i= m_{1/2}$  and in this case one has
$C_3= \sum_{i,j=1,2,3}  \tilde C_3^{ij}$ and  $C_4= \sum_{i=1,2,3} \tilde  C_4^{i}$.  
In Fig. (9) we display the dependence of $\tilde C's$ on the RG scale $Q$.
Here one finds that in addition to $C_1$, 
$\tilde C_3^{11}$ and $\tilde C_3^{22}$ assume negative values which gives
the possibility of new focal curves. We discuss these possibilities in further detail below.
\\

To examine the focal curves and focal surfaces for NuSUGRA, it is useful to define
\beqn
\label{c3g}
C_3^{G}\mhf^2=\tilde{C}_3^{ij}m_im_j,~~~~~C_4^G\mhf =\tilde{C}_4^im_i~.
\eeqn
Further, in order to classify various regions of the  radiative electroweak symmetry breaking  (REWSB) for the 
NuSUGRA   case  it useful to write the REWSB constraint Eq.(\ref{4.1}) in the form
\begin{equation}
\mu^2+\frac{1}{2}M_Z^2=C_1{m_0}^2+C_2{\overline{A}_0}^2+C_3^{(i)}{\overline{m}_i}^2,
\label{REWSB_reduced_form}
\end{equation}
with
\begin{eqnarray}
\overline{A}_0^2 &=& (A_0 + \sum\limits_{i=1}^3 {a_i}m_i)^2,
~~\overline{m}_i 	=
\displaystyle\sum\limits_{j=1}^3 {a_{ij}}m_j ,	
\end{eqnarray}
where $a_i$ and $a_{ij}$ are co-efficients of linear combinations and they are functions of $C_2, 
 \tilde C_3^{ij}$ and $\tilde C_4^i$.\\
\begin{table}[t!]
\begin{center}
\begin{tabular}{|c|c|c|}
\hline
 Focal Curve   & large soft parameters  & small soft parameters \\
\hline 
 HB/FC1 & $m_0-A_0$  & $m_1, m_2, m_3$ \\
 HB/FC2$^{01}$ & $m_0-m_1$  & $A_0, m_2, m_3$ \\
 HB/FC2$^{02}$ & $m_0-m_2$  & $A_0, m_1, m_3$ \\
 HB/FC2$^{03}$ & $m_0-m_3$  & $A_0, m_1, m_2$ \\
 HB/FC3$^{13}$ & $m_1-m_3$  & $m_0, A_0, m_2$ \\
 HB/FC3$^{23}$ & $m_2-m_3$  & $m_0, A_0, m_1$ \\   
 HB/FC4$^{1}$ & $A_0-m_1$  & $m_0, m_2, m_3$ \\   
 HB/FC4$^{2}$ & $A_0-m_2$  & $m_0, m_1, m_3$ \\        
  \hline
\end{tabular}
\label{tab_nuga}
\caption{
Classification of focal curves  in NuSUGRA  models.
Here one has the possibility of several focal curves. The focal curve HB/FC1 is
defined similar to the  mSUGRA case except  that 
$m_1, m_2, m_3$ are all kept fixed. As in mSUGRA  here too 
$m_0$ and $A_0$ can get large while $\mu$ remains fixed. The focal 
curve HB/FC2 splits into three sub cases because of the gaugino non-universalities.
Thus the case HB/FC2$^{01}$ corresponds to the case when  $A_0,m_2,m_3$ are 
kept fixed while $m_0$ and $m_1$ can get large. The focal curves HB/FC2$^{02}$ and
HB/FC$^{03}$ are similarly defined. For the NuSUGRA case  4 new type of 
focal curves arise. These are HB/FC3$^{13}$, HB/FC3$^{23}$, HB/FC4$^{1}$, HB/FC4$^{2}$.
Their definitions are obvious from the table. }
\end{center}
\end{table}

 \begin{figure}[h!]
\begin{center}
\includegraphics[scale=0.20]{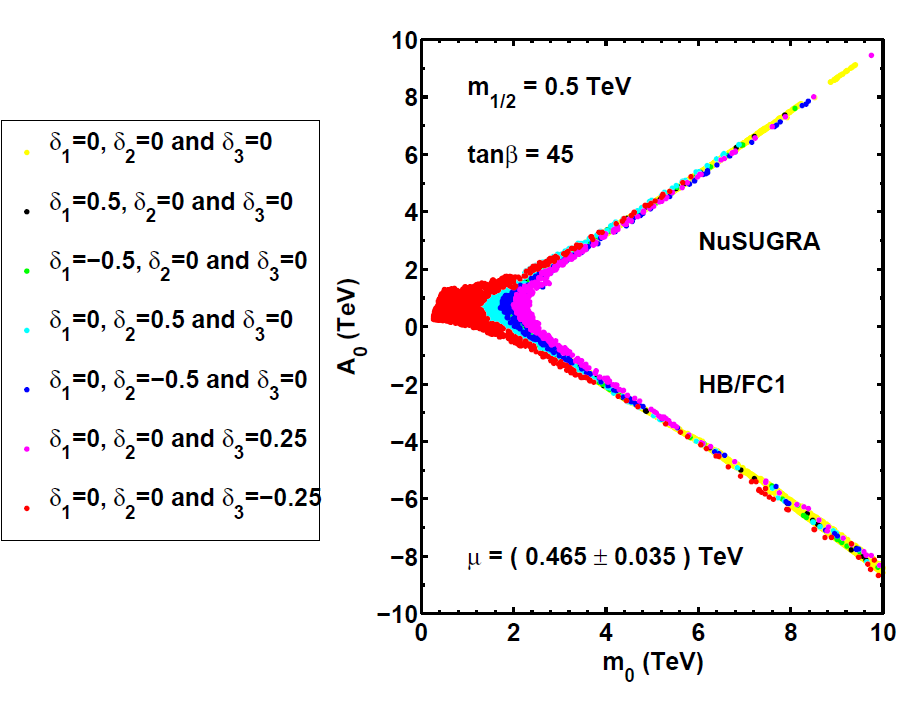}
\includegraphics[scale=0.20]{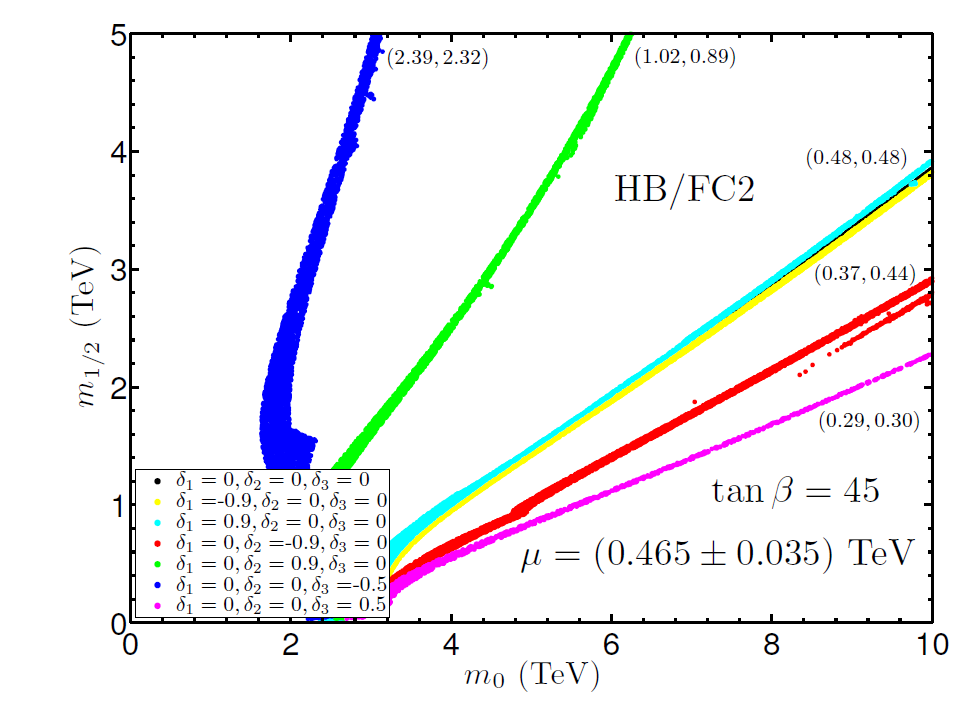}\\
\includegraphics[scale=0.17]{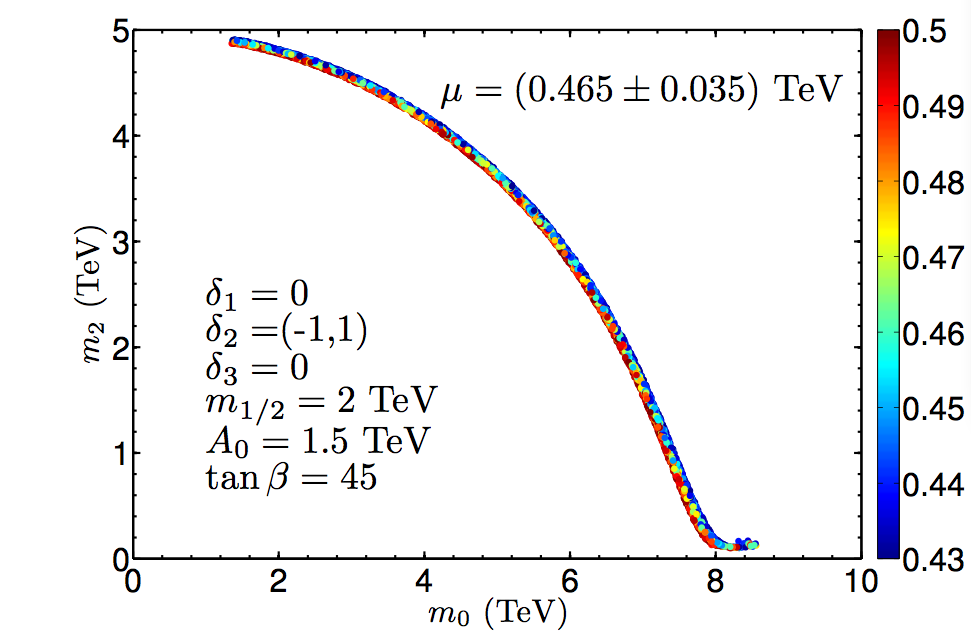}
\includegraphics[scale=0.17]{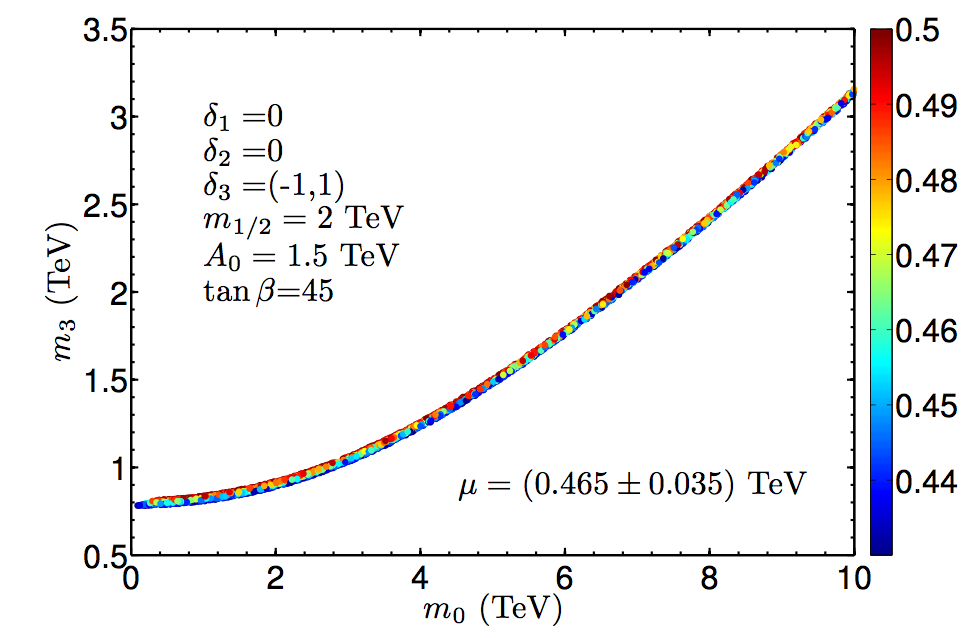}
\includegraphics[scale=0.17]{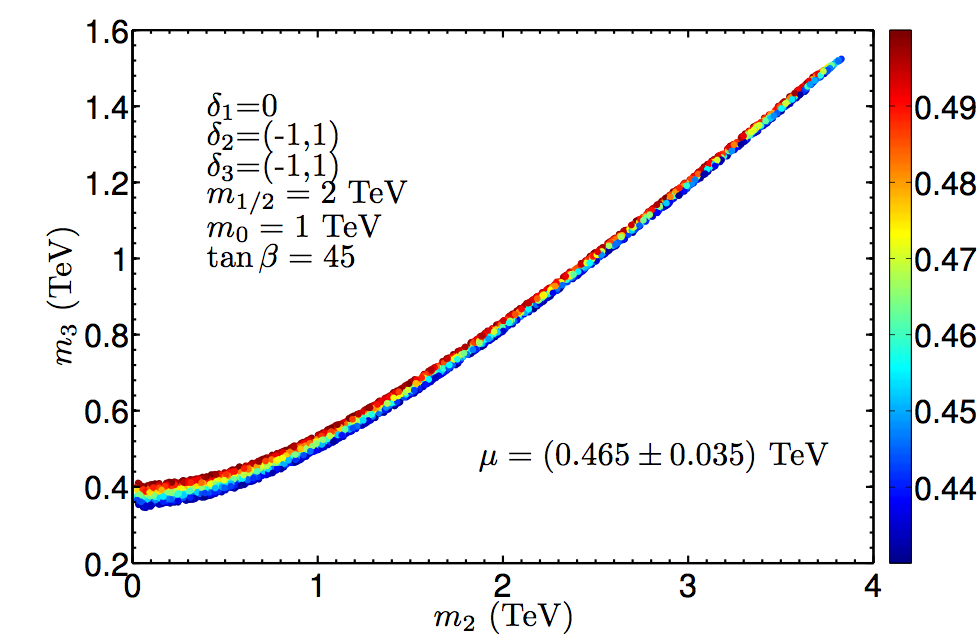}
\caption{
Top left panel:  Exhibition of the Focal~Curve HB/FC1 of Table~II  
with non-universalities in the gaugino sector. Here and in the right panel   $\tan\beta=45$ with $\mu=\left(0.465\pm0.035\right)\TeV$.    
The plot shows that non-universalities in the gaugino sector do not affect
the  asymptotic behavior of $A_0/m_0$ which is unchanged from the mSUGRA case. 
Top right panel: Exhibition of the effect of non-universalities on focal curves FC2.  The analysis shows
that the non-universalites have a very significant effect of FC2 type focal curves. The asymptotic 
form of the FC2 curves with non-universalities fits well with the result of Eq.(\ref{fcnonuni}).
Bottom panels show the three variety of FC2 curves;
 left panel: An exhibition of the Focal~Curve~HB/FC2$^{03}$ in the $m_0-m_3$~plane when $m_1=m_3=m_{1/2}=2\TeV$ and $A_0=1.5\TeV$;
middle panel: A display of the Focal~Curve~HB/FC2$^{02}$  in the $m_0-m_2$~plane  when $m_1=m_3=m_{1/2}=2\TeV$ and $A_0=1.5\TeV$;  
 right  panel: An exhibition of the Focal~Curve~HB/FC3$^{23}$ in the $m_2-m_3$~plane when $m_1=m_{1/2}=2\TeV$, $m_0=1\TeV$ and $\left|A_0/m_0\right| < 0.1$. The model points are colored by $\mu$ value in units of$\TeV$.}
\label{1FC}
\end{center}
\end{figure}

 \begin{figure}[h!]
\begin{center}
\includegraphics[scale=0.25]{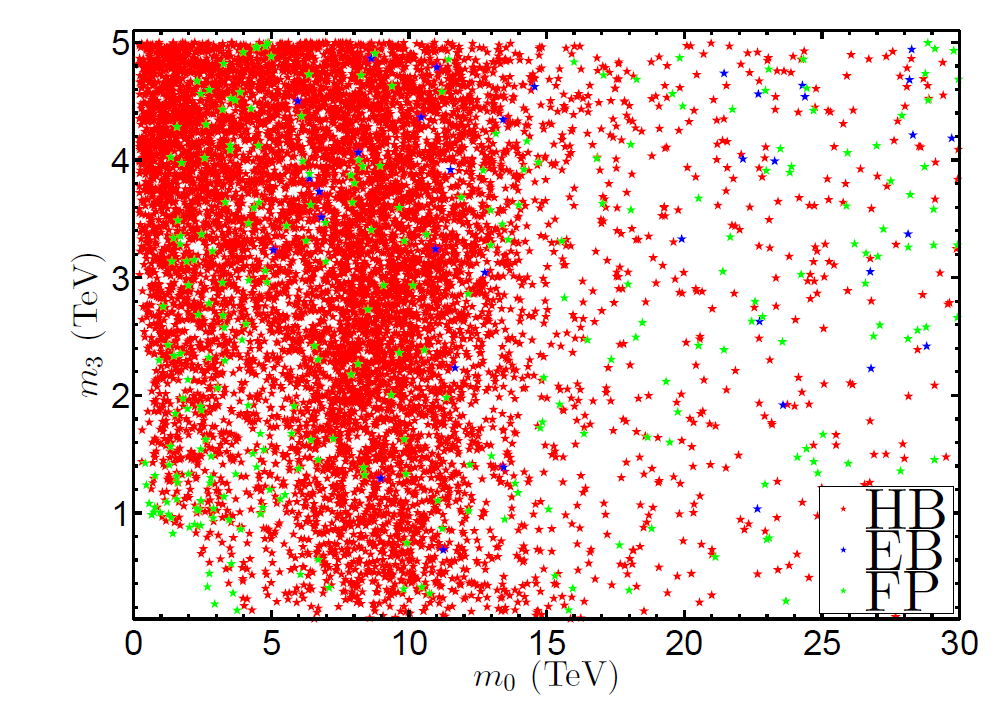}
\caption{\label{nghb}
Exhibition of HB (red),  EB (blue), FP (green)  parameter points for NuSUGRA using the
inputs given in Sec.(\ref{SecEWSB}).
All parameter points satisfy the general constraints along with a $2\sigma$ constraint on the Higgs boson mass. 
As in the supergravity unified models with universal 
boundary conditions here too one finds that  most of the allowed parameter space lies on 
the HB branch while EB and FP regions are highly depleted. }
\end{center}
\end{figure}

A display of the renormalization group evolution of the $C_i$ is given in Fig.~(\ref{Cval}) 
in  Appendix B.  Here we find that in addition to $C_1$, 
the elements 
$\tilde C_3^{11}$ and     $\tilde C_3^{22}$  are   negative,  
which allows for the possibility of new focal curves and focal sufaces over the ones discussed in Sec.(II).
Using the results of Appendix A and B one finds that four types of focal curves arise 
for the NuSUGRA case, FC1-FC4, which are
listed in Table II.  FC1 is defined similar to the case for mSUGRA.  FC2 has three variations:
These are 
HB/FC2$^{01}$ where  $C_1>0, \tilde C_3^{11}<0$ and
 $m_0$ and $m_1$ get large while  $A_0, m_2, m_3$ and $\tan\beta$ 
remain fixed; 
HB/FC2$^{02}$ where $C_1>0, \tilde C_3^{22}<0$  and
 $m_0$ and $m_2$ get large while  $A_0, m_1, m_3$ and $\tan\beta$ 
remain fixed, and  
HB/FC2$^{03}$ where $C_1<0, \tilde C_3^{33}>0$ and $m_0$ and $m_3$ 
get large while  $A_0, m_1, m_2$ and $\tan\beta$ 
remain fixed. It is convenient to use the parametrization of Eq.(\ref{c3g}) to exhibit the
effect of non-universality on focal curves FC2. Thus here one finds that the asymptotic value of
$m_{1/2}/m_0$ for fixed $\mu$ as $A_0$ gets large is affected by non-universality, i.e., one gets
\beqn
\frac{m_{1/2}}{m_0} \to \sqrt{\frac{|C_1|}{C_3^G}}.
\label{fcnonuni}
\eeqn
An illustration of the dependence of ${m_{1/2}}/{m_0}$ on non-universalities  for FC2   will be exhibited shortly. \\

  The focal curves FC3 arise  when two of the gaugino masses get large
while other soft parameters  remain fixed. There are two possibilities here. 
The first one is 
HB/FC3$^{13}$ where $m_1$ and $m_3$ get large while $A_0, m_0, m_2$ and $\tan\beta$ 
remain fixed. 
 This can happen when $C_1>0$ but   $\tilde C_3^{11}$ is negative.
 The second possibility is    
HB/FC3$^{23}$ where  $m_2$ and $m_3$ get large while  $A_0, m_0, m_1$ and $\tan\beta$ 
remain fixed.  This can happen when $C_1>0$ but   $\tilde C_3^{22}$ is negative.
The focal curves FC4 arise when $A_0$ and one of the gaugino masses get large while  the remaining
soft parameters remain fixed. There are two possibilities here. 
The first one is 
HB/FC4$^{1}$ where  $A_0$ and $m_1$ get large while  $ m_0, m_2, m_3$ and $\tan\beta$ 
remain fixed.  This can happen since $C_2>0$ but   $\tilde C_3^{11}$ is negative.
The second possibility is 
 HB/FC4$^{2}$ where  $A_0$ and $m_2$ get large while  $ m_0, m_1, m_3$ and $\tan\beta$ 
remain fixed.  
This can happen when  $C_2>0$ but   $\tilde C_3^{22}$ is negative.
 We note that HB/FC3$^{12}$  does not materialize since $\tilde C_3^{11}$ and 
$\tilde C_3^{22}$ are both negative. Similarly HB/FC4$^{3}$ does not occur since
$C_2$ and $\tilde C_3^{33}$ are both positive.  Further, 
while in principle  HB/FC2$^{03}$
can occur when $C_1<0$ and $\tilde C_3^{33}$ is  positive, the numerical sizes
do not favor appearance of this branch. Thus 
 as shown in the figures in Appendix~\ref{coeff_nug},  $\tilde C_{3}^{aa}$ satisfy $|\tilde C_3^{11} |\ll |\tilde C_3^{22}|\ll  |\tilde C_3^{33}|$, where each step is roughly a factor of $10$. Thus in practice the focal curve HB/FC2$^{03}$ 
 does not materialize.
 Further, for any value of $\tan\beta$, the coefficient $C_1$ begins positive and for $\tan\beta\lesssim 5$ it never becomes negative (for $Q\lesssim 10\TeV$).  
Because of the above  additional possibilities such as HB/FC3$^{12}$ etc are  not realized.  \\

\begin{figure}[h!]
\begin{center}
\includegraphics[scale=0.3]{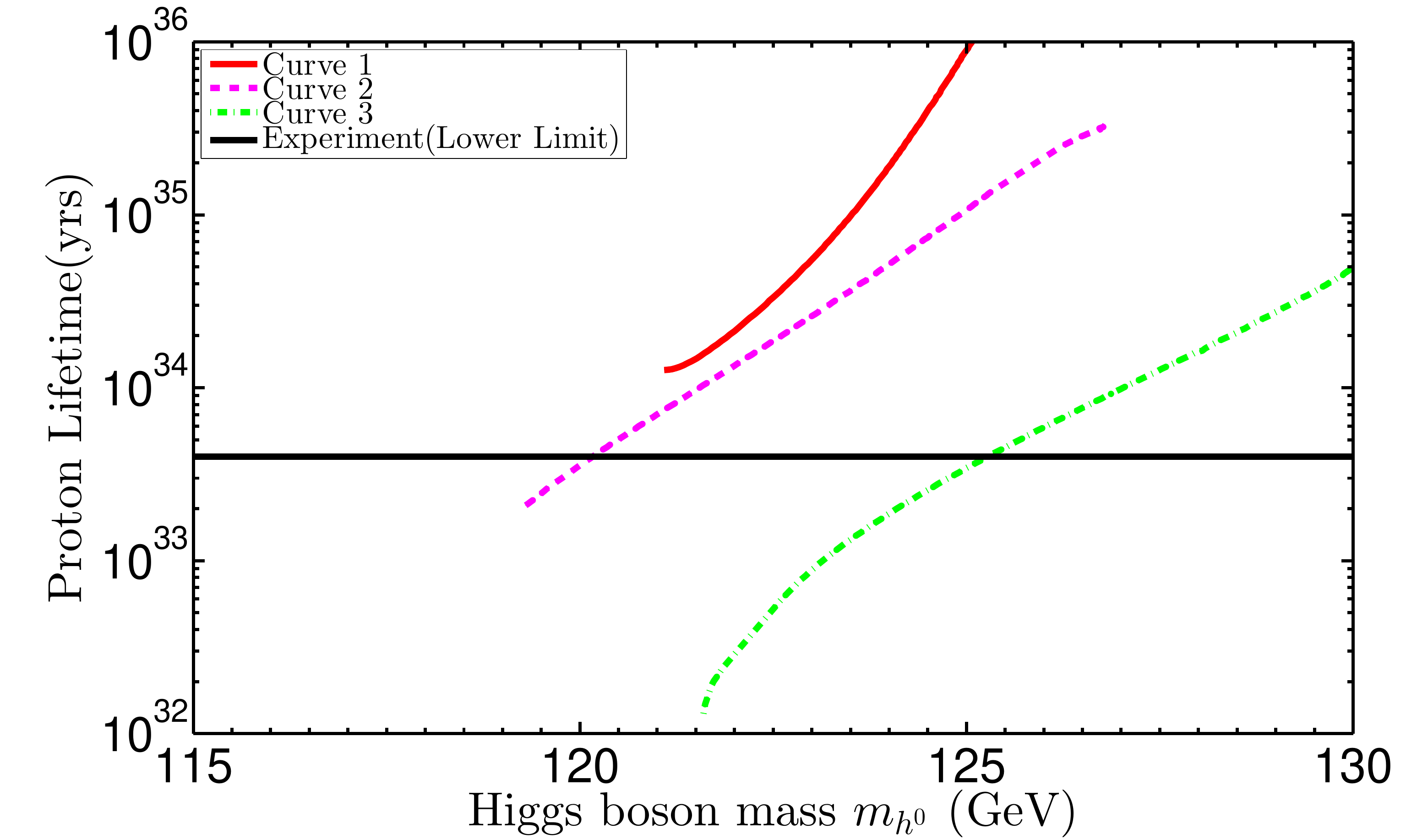}
\caption{\label{ngplife}
An exhibition of the dependence of the proton lifetime for the decay mode $p\to \bar \nu K^+$ 
as a function of the Higgs boson mass for  NuSUGRA. 
Parameters for the curves labelled 1-3 in the legend are as follows:
Curve 1:  $m_1=  4230\GeV$, $m_2= 843\GeV$,  $m_3 =3285\GeV$, $A_0 = -27545\GeV$, $\tan\beta=  5.3$ 
while $m_0$ varies and $M_{H_3}^{eff}/M_G=50$ here and for other curves;
Curve 2: $m_1= 4794\GeV$, $m_2= 3837\GeV$, $m_3= 3856\GeV$, $A_0 / m_0 = 0.842$, $\tan\beta= 7.0$
while $m_0$ and $A_0$ vary;
and
Curve 3: $m_1=3894\GeV$, $m_2=  1056\GeV$, $m_3=  2345\GeV$, $A_0 / m_0 = 2.199$, $\tan\beta=  55.2$
while $m_0$ and $A_0$ vary.
As for the case of the supergravity unified models with universal boundary conditions, 
here too one finds that the proton lifetime is a very sensitive function of the Higgs boson mass. 
}
\end{center}
\end{figure}
\begin{figure}[h!]
\begin{center}
\includegraphics[scale=0.18]{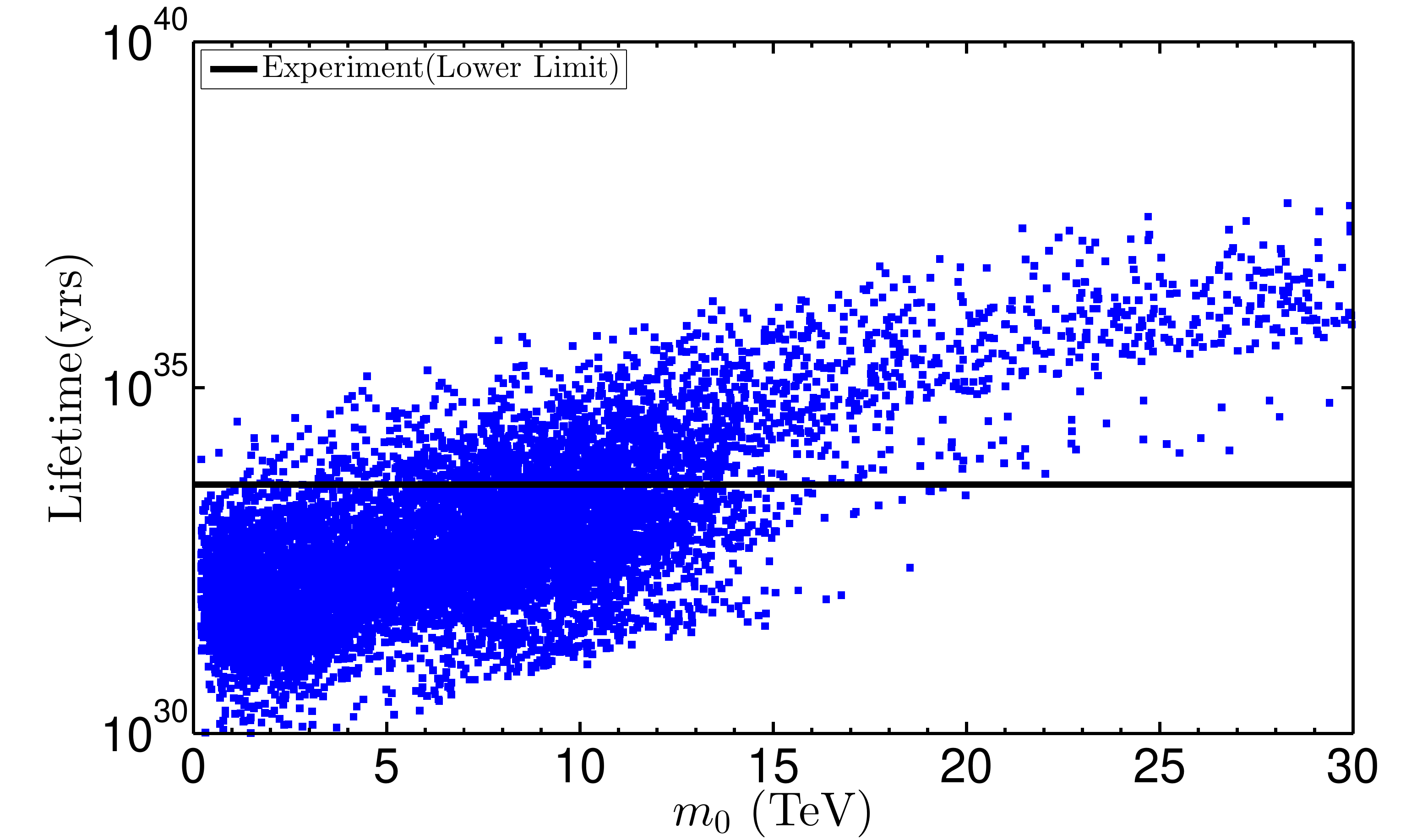}
\includegraphics[scale=0.18]{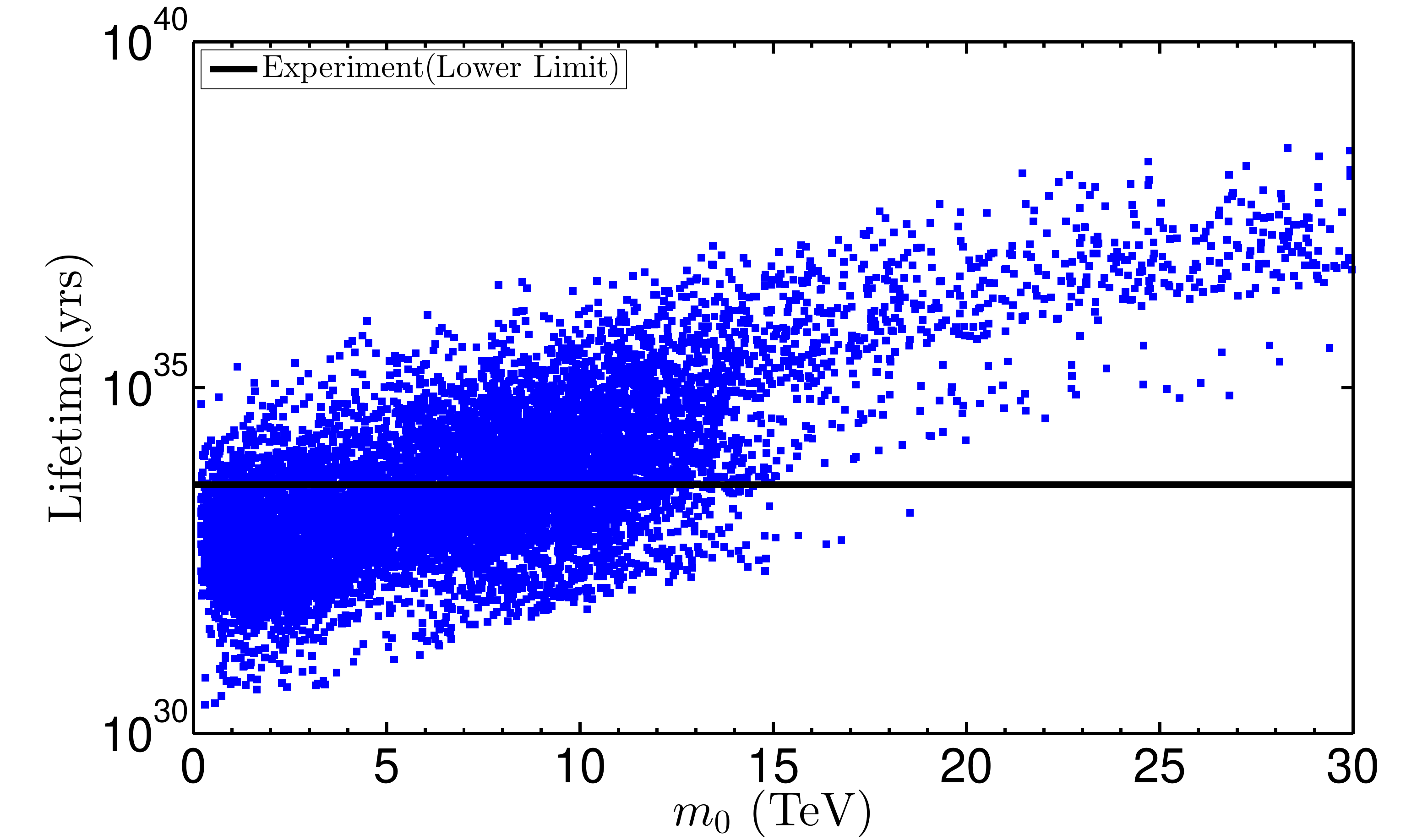}
\includegraphics[scale=0.18]{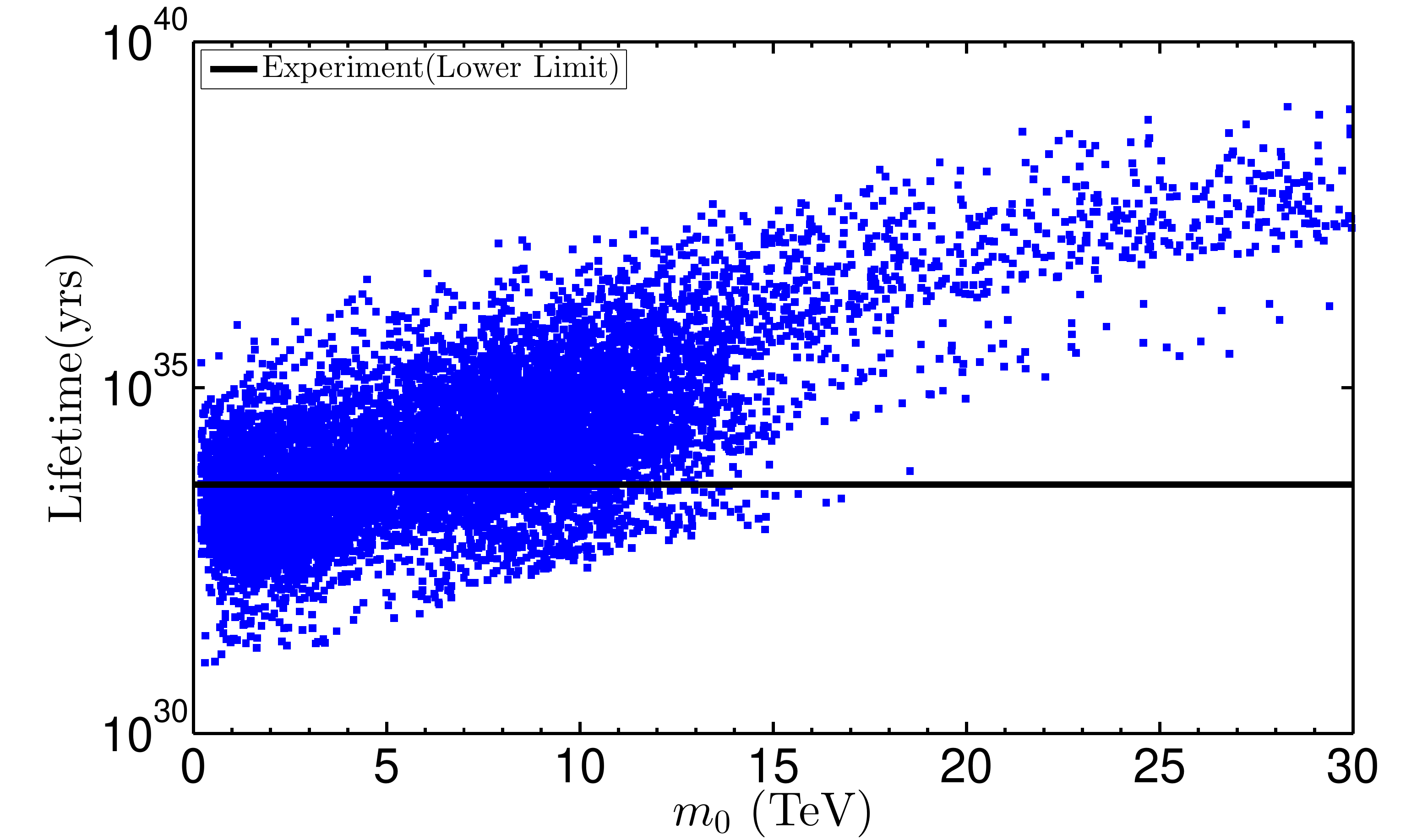}
\caption{\label{ltng}
Left panel: The proton lifetime  shown as blue squares over the allowed parameter space in NuSUGRA models
which  pass the  general constraints along with a $2\sigma$ constraint for the Higgs boson mass as 
discussed in the text when  $M_{H_3}^{eff}/M_G=10$. 
Middle panel: Same as the left panel except that  $M_{H_3}^{eff}/M_G=25$. 
Right panel: Same as the left panel except that  $M_{H_3}^{eff}/M_G=50$. 
The current experimental lower limit on the proton lifetime is shown as a black horizontal line.
The analysis given here is consistent with the Higgs boson mass within a   $2\sigma$  range.}
\end{center}
\end{figure}
 
 For NuSUGRA we give a numerical illustration of some of the focal curves in Fig.~(\ref{1FC}).
The left panel of the top row in Fig.~(\ref{1FC}) gives an analysis of the Focal Curve FC1 in the $m_0-A_0$ plane.
Here one finds that $m_0$ and $A_0$ can get as large as 10 TeV while $\mu$ lies in the range
$(0.465\pm 0.035)$ TeV when $\tan\beta =45$ and $m_{1/2}=0.5$ TeV. 
We note that
 the ratio $A_0/m_0$ asymptotes to the same value irrespective of the non-universalities. 
 A similar analysis for FC2 is given in the right panel of Fig.~(\ref{1FC}) in the $m_0-m_{1/2}$ plane
 for  $\tan\beta=45$.    Again a variety of 
 non-universalities are discussed. One finds  that while $m_0$ and $m_{1/2}$ can get very large, i.e.,
 as large as 10 TeV for $m_0$ and 5 TeV for $m_{1/2}$, one still has a small $\mu$, i.e., a $\mu$ 
 range $(0.465\pm 0.035)$ TeV. 
 An analysis for FC3 is given in the three panels of the bottom row in  Fig.~(\ref{1FC}). 
 The left panel gives a display of the focal curve FC3$^{02}$ in the $m_0-m_2$ plane 
 for the case when $\tan\beta=45$, $A_0=1.5$ TeV, $m_{1/2}= 2 $TeV and $\delta_1= 0 = \delta_3$
 and $\delta_2$ lies in the range $(-1,1)$. One finds that $\mu$ lies in the narrow range 
 $(0.465\pm 0.035)$ TeV.  A very similar analysis in the $m_0-m_3$ plane is given in the 
 middle panel in Fig.~(\ref{1FC}) where $\delta_1=0=\delta_2$ and $\delta_3$
 lies in the range $(-1,1)$ while all other parameters are as in the left panel. This is the focal curve
 FC3$^{03}$.   Finally the right panel gives an analysis of the focal curve FC3$^{23}$  in the ${m}_2
 - {m}_3$ plane for the case when $m_0=1$ TeV, $m_{1/2}=2$ TeV, $\tan\beta=45$ and 
 $\delta_0=0$, $\delta_2 =(-1,1)$, and $\delta_3=(-1,1)$.  Here again one finds that 
 $\mu$ lies in the range $(0.465\pm 0.0350)$ TeV while $m_2,m_3$ get large. 
 From a convolution of the focal curves one can generate focal surfaces where 
 more than  two soft parameters
 can vary while $\mu$ remains fixed.  \\

In Fig.(\ref{nghb}) we display the nature of radiative breaking  of the electroweak symmetry
for all the model points
within the allowed ranges of the parameter space for NuSUGRA.
The points in red are those that lie on HB, the points in blue lie on EB and the points in green
lie in the FP region as defined by Eqs.(\ref{1.6}) and (\ref{1.7}). As in the mSUGRA case here too
one finds that most of the parameter points lie on HB and only a small fraction lie on EB and FP. 
 In Fig.(6) we give an analysis of the sensitivity of the proton lifetime to the Higgs boson
 mass for NuSUGRA. As in the mSUGRA case here too one finds that
 the proton lifetime is very sensitive to the Higgs boson mass with the proton lifetime changing by
 over two orders of magnitude with a shift in the Higgs boson mass in the range of 5-10 GeV. 
 In Fig.(\ref{ltng}) an analysis of the proton lifetime for the mode $p\to \bar \nu K^+$ is given 
 over the allowed parameter space of NuSUGRA within
 the assumed limits. The figure shows the dispersion in the proton lifetime as all the parameter
 points are varied but does show the general trend that $p\to \bar \nu K^+$ lifetime increases 
 with a larger SUSY scale.

 \begin{figure}[h!]
\begin{center}
\includegraphics[scale=0.25]{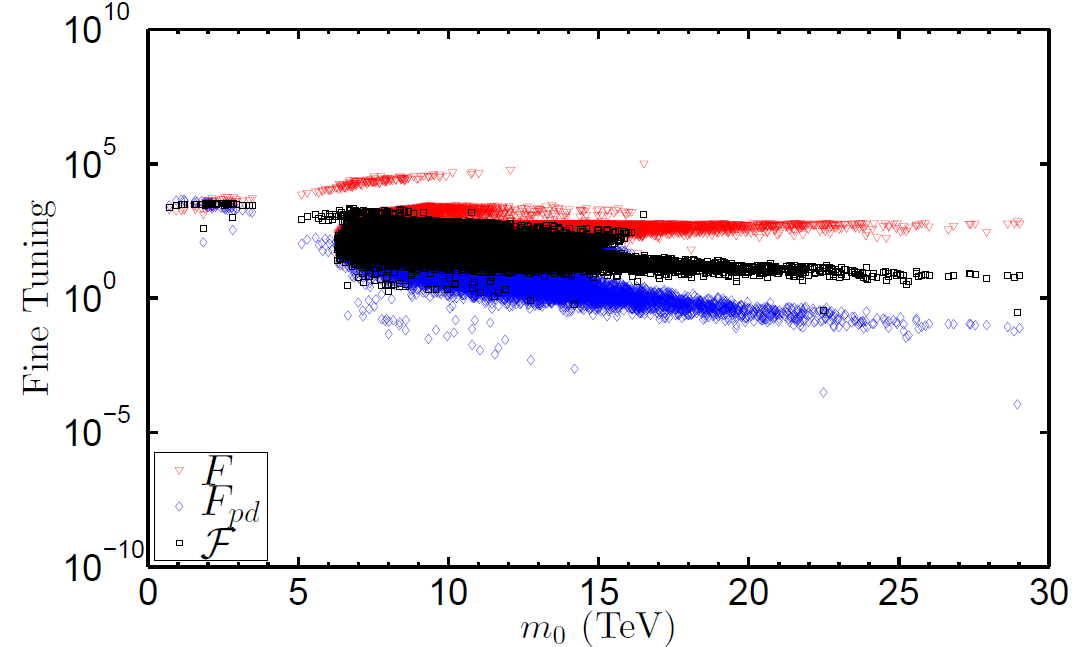}
\includegraphics[scale=0.25]{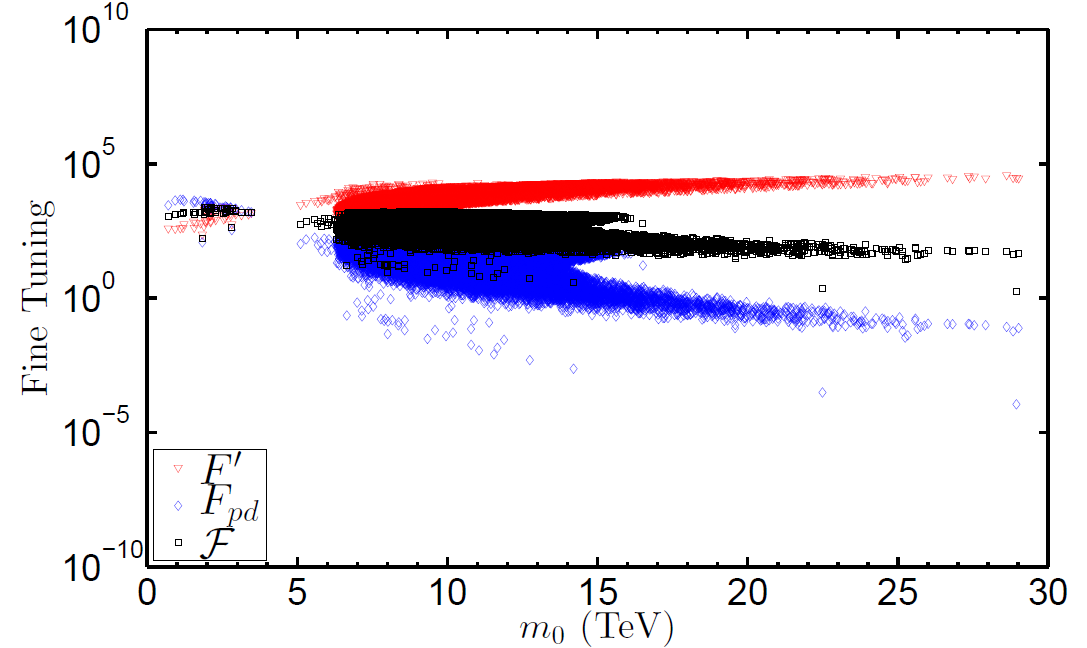}\\
\includegraphics[scale=0.25]{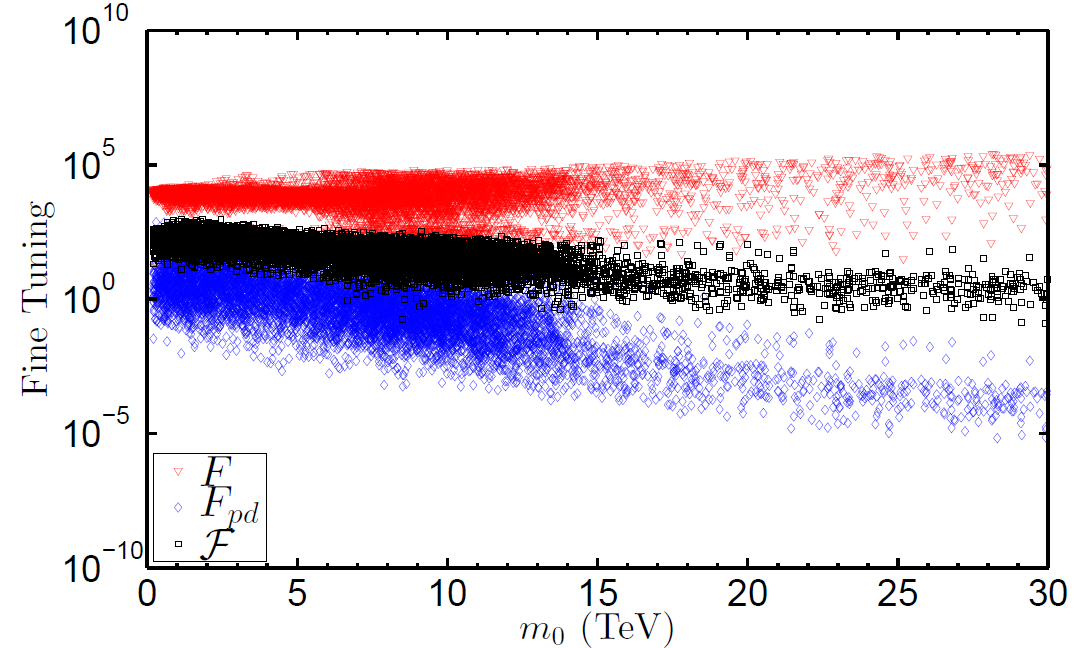}
\includegraphics[scale=0.25]{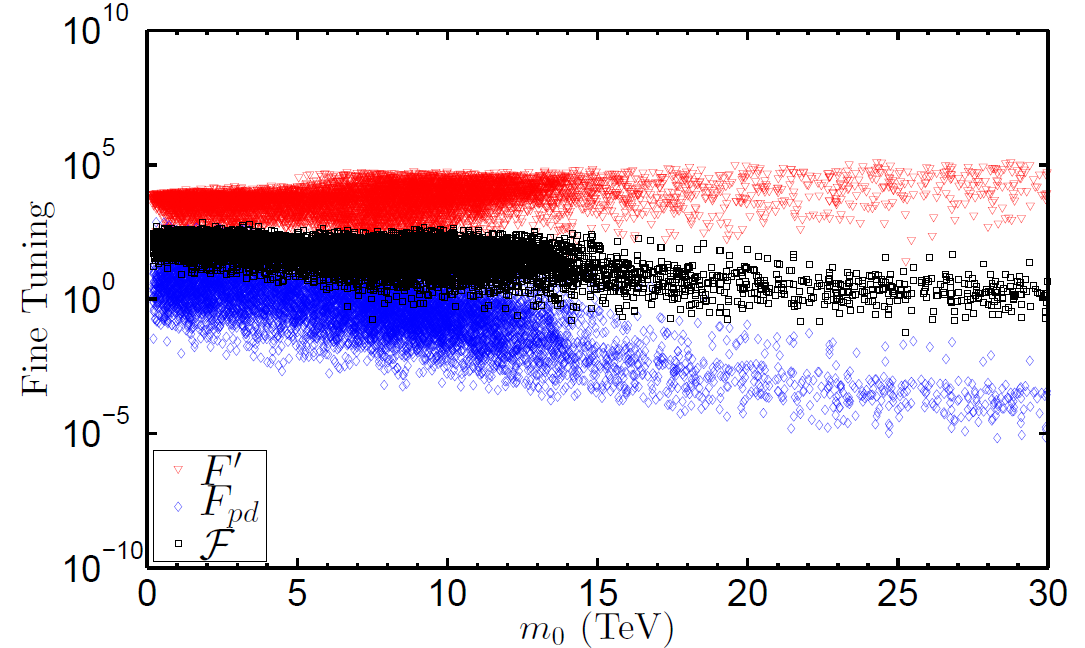}\\
\includegraphics[scale=0.25]{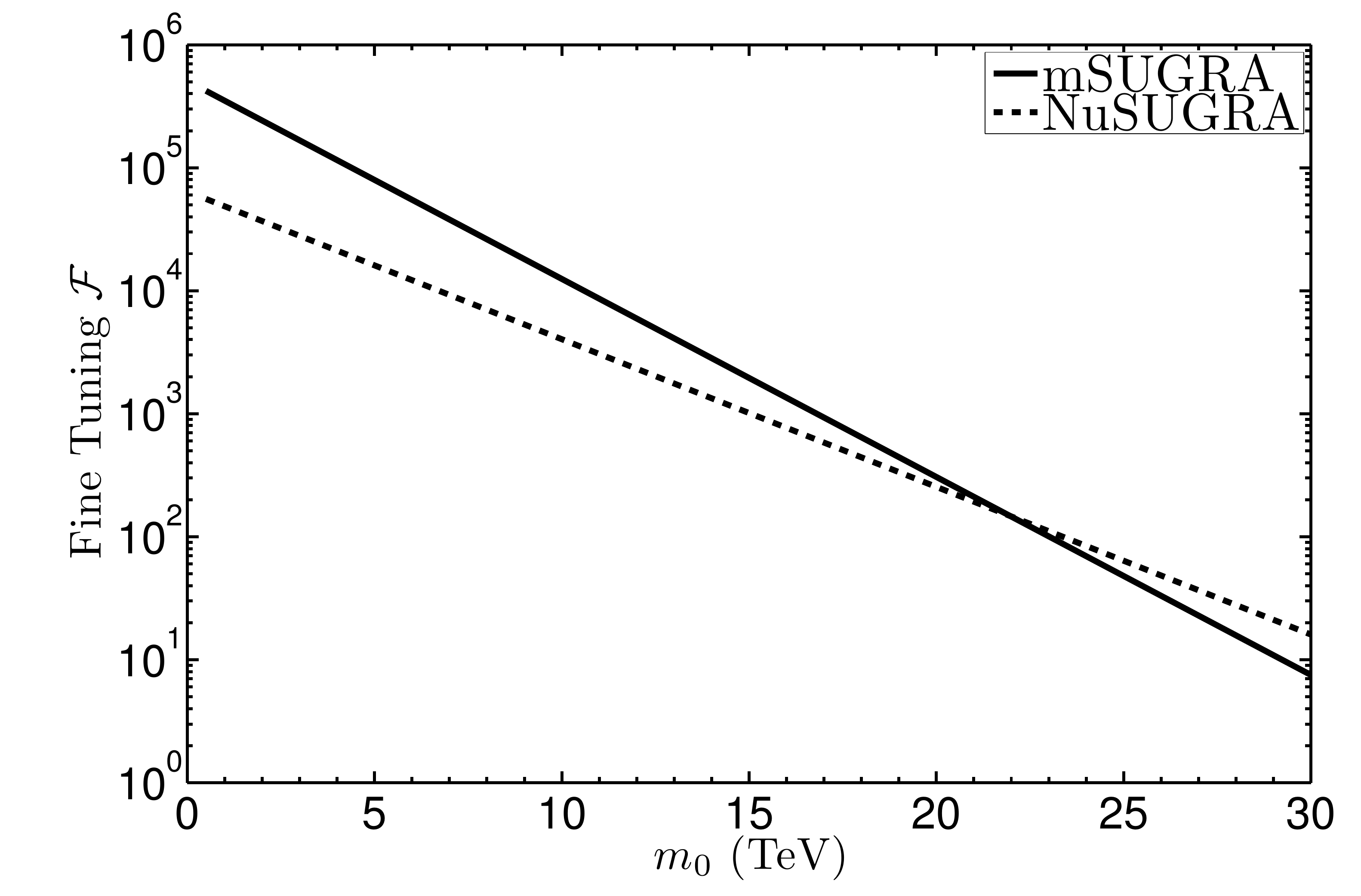}
\caption{\label{finetuning}
A display of the fine tuning as defined by Eqs.(\ref{5.2}-\ref{5.5}) vs the scalar 
mass $m_0$ when   $M_{H_3}^{eff}/M_G=50$.
The upper two panels are for mSUGRA and the middle two panels are for the NuSUGRA case. 
The left panels are when we use the fine tuning of 
Eq.(\ref{5.2}) and the right panels are when we use the fine tuning of Eq.(\ref{5.3})
for the electroweak sector. 
The red points are the fine tunings values for the REWSB sector, the blue points for 
$\tau(p\to \bar \nu K^+)$, and the black points are the averages of the red and the blue points. 
In the bottom panel the combined  fine tuning as a function
of $m_0$ is given for  mSUGRA  (sold line) and for NuSUGRA (dashed line).  Here we have taken the average of the
left and right panels and drawn smooth curves showing the rapid decrease of the fine tunings
as $m_0$ increases. 
} 
\end{center}
\end{figure}

\section{Naturalness\label{ft}}
 
 The criteria used for quantifying what is naturalness are rather subjective and
various variants abound
see, e.g.,\cite{bg,Ciafaloni:1996zh,Bhattacharyya:1996dw,Chan:1997bi,ACR,Kane:1998im,Chankowski:1998xv,Casas:2003jx,Martin:2010kk,ArkaniHamed:2012kq,Ghilencea:2012gz,Jaeckel:2012aq,Baer:2012cf,Baer:2012mv}. 
Here we discuss the fine tuning within a GUT framework including both radiative breaking of the 
electroweak symmetry and proton stability. First we discuss fine tuning for radiative breaking of the 
 electroweak symmetry which is governed by the breaking condition Eq.(\ref{1.1}).
If one views $M_Z^2$ as arising from the cancellation between $\mu^2$ term and the remainder on the right hand
side, it leads to a fine tuning~ \cite{Chan:1997bi}
\begin{eqnarray}
F\simeq \frac{4\mu^2}{M^2_Z}.
\label{5.2}
\eeqn
An alternate criteria for fine tuning 
is given by the condition\cite{bg} 
$F'_a= (a/f(a)) f'(a)$
where $a$ is the sensitive parameter on which the function $f(a)$ depends. Using $f(a)= M_Z^2$ and the sensitive parameter as $m_{H_u}^2$ one 
finds  another fine tuning measure
 \beqn
F'\simeq \frac{2\vert\mhu\vert}{M^2_Z}.
\label{5.3}
\eeqn
We will use both $F$ and $F'$ in the analysis for comparison. For proton decay we will use  a measure of fine tuning 
defined by 
\beqn
F_{pd} = \frac{4\times 10^{33} {\rm yr}}{\tau(p\to \bar \nu K^+){\rm yr}}.
\label{5.4}
\eeqn
This measure gives the amount of fine tuning needed in the theory parameters to enhance the lifetime so 
that the theoretical prediction is brought 
 just above the current experimental lower limit. 
 If we use the very crude approximation on the proton lifetime, i.e., 
 $\tau(p\to \bar \nu K^+)\simeq C \cdot ({m_{\tilde \chi^{\pm}}}/m_{\tilde q}^2 M_{H_3}^{eff})^{-2}$ and use
  $m_{\tilde q}^2$ or  $m_{\chi^{\pm}}$ as the sensitive parameters, 
 we have $F'_{m_{\chi^{\pm}}}=F'_{m_{\tilde q}^2}=2F_{pd}$. 
 Thus the two ways of defining the fine tuning differ only by a small numerical factor.
 It is also useful to define a composite fine tuning by the geometric mean of the individual ones, i.e., 

\be
{\cal F}=\left(\prod_{i=i}^n F_i\right)^{\frac{1}{n}}.
\label{5.5}
\ee
Here our view point is similar to that of~\cite{Jaeckel:2012aq}
(for a related work see~\cite{McKeen:2013dma}).
For our case $n=2$ consisting of the fine tuning in the radiative electroweak symmetry breaking sector and the 
 fine tuning needed to control proton decay from dimension five operators.
  An analysis of the fine tunings as a function of $m_0$ is given in Fig.(\ref{finetuning}) 
 where the upper panels give the analysis for the case of mSUGRA and the middle panels
give  the  analysis for NuSUGRA, and  where the left panels give the analysis
using Eq.(\ref{5.2}) and the right panels give the analysis using Eq.(\ref{5.3}). 
The red points are the fine tunings
for radiative electroweak symmetry breaking.
The blue points give the fine tuning needed in the theory prediction of $\tau(p\to \bar \nu K^+)$ to 
bring the lifetime prediction just above the experimental lower limit,
 and the black points 
correspond to the composite fine tuning as defined by Eq.(\ref{5.5}). One finds that
typically there is a preference for larger values of $m_0$ for the combined
fine tuning including fine tuning from the electroweak sector and the fine tuning 
needed from proton stability. 
This result is more explicitly exhibited in the bottom panel of Fig.(\ref{finetuning}) 
which shows fine tuning prefers regions of larger $m_0$ 
  when the electroweak symmetry breaking
and proton stability criteria are combined. A similar conclusion was arrived at in 
the work of  ~\cite{Jaeckel:2012aq} which combined the electroweak symmetry
breaking, FCNC and CP violation criteria.

\section{Conclusions\label{SecConclu}}
The high mass of the Higgs boson discovered recently requires a large loop
correction to its mass which points to the possibility that the overall weak scale
of supersymmetry may lie in the several TeV region and could even be as large
as tens of TeV. If the scalar masses are that large they would help
resolve  one of the serious problems of supersymmetric grand unification
related to proton decay. Thus proton decay from lepton and baryon number
violating dimension five operators often leads to proton lifetimes which 
fall below the  current experimental limits.
In this work we show that the proton lifetime is a very sensitive 
function of the Higgs boson mass in a unified theory. Thus  a few GeV upwards shift in 
the Higgs boson can result in orders of magnitude suppression of the
proton decay from baryon and lepton number violating dimension five operators
and a corresponding enhancement of the proton lifetime. 
The analysis is first done for the mSUGRA model and then extended to 
NuSUGRA.  
Here we also analyse the allowed parameter space in terms of which 
branch of the radiative breaking of the electroweak symmetry the parameters
lie, i.e., whether on the Ellipsoidal Branch, the Hyperbolic Branch or the Focal Point region.
The analysis presented in this work shows that 
under the current experimental constraints including those from LEP, Tevatron, LHC, 
FCNC and the Planck data~\cite{Boehm:2013qva}
 one finds that most of the parameter 
points of mSUGRA and of NuSUGRA models 
lie on the Hyperbolic Branch with only a very small fraction 
lying on the Ellipsoidal Branch or in the Focal Point region.  
We also discuss issues of naturalness and fine tuning and show
that the composite fine tuning including fine tuning from
 the electroweak sector and from the stability of the proton 
points to high scalar masses.  However, some of  the gauginos can be
light with their masses mostly limited  by their lower experimental 
limits. These include the light chargino, the lightest neutralino,
and the second lightest neutralino and the gluino.  These should be
accessible with increased energy and luminosity at the 
next round of experiment at the LHC. Regarding proton decay
discovery of  the supersymmetric mode $p\to \bar \nu K^+$  
is over due and this mode continues to be the most likely  candidate
to
be discovered first in the next generation of proton decay experiments.
\\

\section*{Acknowledgments}
Collaboration in the initial stages of this work with Sujeet Akula and Gregory Peim is acknowledged.
ML also thanks them for help with numerical analysis in this work.
This  research is  supported in
part by the U.S. National Science Foundation (NSF) grants 
PHY-0757959 and  PHY-070467 and DOE  NERSC grant DE-AC02-05CH11231.
 \appendix
\begin{appendix}
\section{The relation of $C_1(Q)$ to $m_{H_2}$ \label{SecEBHB}}
Here we establish the relation between the $C_1(Q)$ and $m_{H_2}$. The RG evolution connects $m_{H_2}$  
and on the third generation masses 
$m_U$ and
$m_Q$ and one has 

\beqn
\frac{d}{dt}\left[\begin{array}{ccc}
         m^2_{H_2}           \\
         m_U^2    \\
         m_Q^2        
	\end{array}\right] 
= - Y_t  \left[\begin{array}{ccc}
        3      &  3 & 3           \\
        2      &   2 &   2  \\
        1 &  1 & 1 \\
	\end{array}\right]
\left[\begin{array}{ccc}
        m^2_{H_2}           \\
         m_U^2    \\
         m_Q^2        
	\end{array}\right]
		 -Y_t A_t^2 \left[\begin{array}{ccc}
         3           \\
         2   \\
         1        
	\end{array}\right]	
	+ 
		  \left[\begin{array}{ccc}
 3\tilde\alpha_2m_2^2+
\tilde\alpha_1m_1^2                   \\
 \frac{16}{3}\tilde\alpha_3m_3^2
+ \frac{16}{9} \tilde\alpha_1m_1^2           \\
\frac{16}{3}\tilde\alpha_3m_3^2
+ 3\tilde\alpha_2m_2^2+
\frac{1}{9} 
\tilde\alpha_1m_1^2                
	\end{array}\right].	
\eeqn

Here $Y_t=h_t^2/(4\pi^2)$ where $h_t$ is the top Yukawa coupling and $A_t$ is the trilinear 
coupling in the top sector. The  above equations with  universal boundary conditions at the GUT scale allow 
 a homogeneous solution satisfying~\cite{Feng:1999mn} 

\be
\left[\begin{array}{c}
        \delta m^2_{H_2}           \\
        \delta m_U^2    \\
        \delta m_Q^2        
	\end{array}\right] 
= \frac{m_0^2}{2}   \left[\begin{array}{c}
        3J(t)-1          \\
        2 J(t)~~~~~ \\
        J (t) +1 
	\end{array}\right]~,
	\ee
where $J$ is an integration factor defined by 
\be
J(t) \equiv  \exp\left[-6\displaystyle\int_0^{t} Y_t(t') dt'\right]~.
\label{j1}
\ee
As $Q \to M_G$, one has $J(t)\to 1$ and the universality of the masses is recovered at the GUT scale.
In \cite{Akula:2011jx}
a  connection was established between $C_1(Q)$ and $\delta m_{H_2}$ which we now
illustrate. Thus  $Y(t)$ at the one loop level satisfies the equation
\be
\frac{dY_t}{dt} = \left(\frac{16}{3} \tilde \alpha_3 + 3 \tilde\alpha_3+\frac{13}{9} \tilde \alpha_1\right) Y_t - 6 Y_t^2~,
\ee
 and one finds 
\be
Y_t(t) = \frac{Y(0) E(t)}{1 + 6 Y(0) F(t)}~,
\label{aj2}
\ee
where $F(t)$ and $E(t)$ are defined after Eq.~\eqref{1.4}. It is then easy to see that $J(t) = D_0(t)$,
where $D_0(t)$ is defined by Eq.~\eqref{1.4}. 
Thus   $\delta m^2_{H_2}$ takes the form 
\be
\delta\overline{m}_{H_2}^2 \equiv \frac{\delta m^2_{H_2}}{m_0^2}   = \frac{1}{2} \left(3D_0-1\right)~,
  \label{j3}
\ee
and $C_1$ can be expressed in terms of $\delta\overline{ m}_{H_2}^2$
\be
C_1= \frac{1}{\tan^2\beta-1}\left(1- \delta\overline{ m}_{H_2}^2\tan^2\beta\right)  \simeq  - \delta\overline{ m}_{H_2}^2    ~.
\label{j4}
\ee
$C_1=0$ was defined as the Focal Point in \cite{Akula:2011jx}. At the Focal Point
  $\mu^2$ essentially becomes independent of $m_0$. For $\tan\beta>>1$,
$C_1\simeq  - \delta\overline{ m}_{H_2}^2$ and  the vanishing of $C_1$ implies vanishing of $\delta\overline{ m}_{H_2}^2$
which is defined to be the Focus Point.  Thus the Focal Point defined by $C_1=0$ 
is just the boundary point between
EB defined by $C_1>0$ and HB define by $C_1<0$. For the NuSUGRA models all solution where
some of the soft parameters can get large while $\mu^2$ remains fixed lie on HB. This can happen when some of the 
$C_i$ other than $C_1$ turn negative, as discussed 
in the Appendix below.

\section{Analysis of $C's$ for Models with  Non-universalities in the Gaugino Sector \label{coeff_nug}}

The presence of non-universalities in the gaugino sector affects the co-efficients $C_i$ and 
in this Appendix we give a computation for these by inclusion of non-universalities in the 
$SU(3)_C$, $SU(2)_L$ and $U(1)$ gaugino sectors. 
We begin with the radiative electroweak symmetry breaking with the inclusion of non-universalities in the
gaugino sector. We have 
\begin{equation}
\label{musqr}
\mu^2=\frac{(\mH{1}-\mH{2}\tbeta^2)}{(\tbeta^2-1)}-\frac{1}{2}M_Z^2+\Delta\mu^2,
\end{equation}
with 
\begin{eqnarray}\label{mHs}
\mH{1} &=& m_0^2+\left(\frac{3}{10}\ft_{1}+\frac{3}{2}\ft_{2}\right), \\
\label{mh1}
\mH{2} &=& \tilde{e}(t)+A_0\tilde{f}(t)+m_0^2h(t)-A_0^2k(t),
\label{mh2}
\end{eqnarray}
and $\tilde f_i(t)$ is defined by $\tilde{f}_i(t)=Z_i^f m_i^2$ where 
\begin{equation}\label{Zfi}
Z_i^f=\frac{1}{\beta_i} \left(1-\frac{1}{(1+\beta_{i}t)^2}\right) {\tilde{\alpha}_i(0)}.
\end{equation}
It is useful to introduce a column vector $\vec{m}^T= (m_1, m_2, m_3)$ and
 a matrix $\MmH{1}$ such that
$\mH{1} = \vec{m}^\mathsf{T}\cdot \MmH1 \cdot\vec{m}=\left(\MmH{1}\right)_{ij}m_im_j$ 
where $M_{H_1}$ is given by 
\begin{equation}
\MmH{1}=
\begin{pmatrix}
\dpsty\frac{3}{10}\Zf{1} & 0 & 0	\\
0 & \dpsty\frac{3}{2}\Zf{2} & 0	\\
0 & 0 & 0 \\
\end{pmatrix}.
\end{equation}
Thus  we have
\begin{equation}\label{mH1}
m_{H_1}^2=m_0^2+\left(\MmH{1}\right)_{ij}m_im_j.
\end{equation}
The above exhibits the gaugino mass dependence of $m_{H_1}^2$ explicitly. 
Now let us look at $m_{H_2}^2$ given by Eq.(\ref{mh2}) and  write it in a form which
 exhibits the gaugino mass dependence explicitly. Now $m_{H_1}^2$ contains 
the functions $\tilde e(t)$ and  $\tilde f(t)$ which are given as 
\begin{equation}\label{et}
\et=\frac{3}{2} \left[ \frac{\tilde{G}_1+Y_0\tilde{G}_2}{D(t)} + \frac{(\tilde{H}_2+6Y_0\tilde{H}_4)^2}{3D(t)^2} + \tilde{H}_8 \right],
~\ft=-\frac{6Y_0\tilde{H}_3(t)}{D(t)^2},
\end{equation}
where $\tilde{H_i}(t)$  are defined by 
\begin{eqnarray}
\label{Ht2}
\Ht{2} =  \frac{13}{15}\tilde{h}_1(t) + 3\tilde{h}_2(t) + \frac{16}{3}\tilde{h}_3(t),
~~~\Ht{3} & = &\dint E(t')\tilde{H}_2(t')dt' ,\\ 
\label{Ht4}
\Ht{4} = F(t)\tilde{H}_2(t)-\tilde{H}_3(t), 
~~\Ht{5} &=&  \left( -\frac{22}{15}\tilde{f}_1(t) + 6\tilde{f}_2(t) - \frac{16}{3}\tilde{f}_3(t) \right) ,\\ 
\Ht{6} = \dint E(t')\tilde{H}_2(t')^2 dt', 
\label{Ht8}
~~\Ht{8} &=& \atG  \left( -\frac{8}{3}\tilde{f}_1(t) + \tilde{f}_2(t) - \frac{1}{3}\tilde{f}_3(t) \right) ,
\end{eqnarray}
and $\tilde h_i$ are defined by $\tilde{h}_i \equiv Z_i^h m_i$
with 
\begin{equation} \label{Zhi}
Z_i^h=\frac{t}{1+\beta_i t}\tilde{\alpha}_i(0).
\end{equation}
$\tilde{H}_2(t)$ then takes the form 
\begin{equation}
\Ht{2}\equiv\vMHt{2}\cdot\vec{m},
\end{equation}
where $\vMHt{2}$ is a row vector 
\label{vMHt2}
\begin{equation}
\vMHt{2}=\left(\frac{13}{15}\Zh{1}, 3\Zh{2}, \frac{16}{3}\Zh{3}\right).  \\
\end{equation}
Similarly, we may write all the $M_{\tilde{H}_i}(t)$ in matrix or vector forms so that 
$
\Ht{3}=\vMHt{3}\cdot\vec{m},
~~\Ht{4}=\vMHt{4}\cdot\vec{m},
~~\Ht{5}=\vec{m}^\mathsf{T}\cdot\MHt{5}\cdot\vec{m},
~~\Ht{6}=\left(\vMHt{2}\cdot\vec{m}\right)^2.
~~\Ht{8}=\vec{m}^\mathsf{T}\cdot\MHt{8}\cdot\vec{m}
$
where the matrices 
$\vMHt{3}$ etc are given by 
\begin{equation}
\label{vMHt3}
\vMHt{3}=\dint E(t')\vMHt{2}(t')dt',	\nonumber\\
~~\vMHt{4}=\vMHt{2}(t)\dint E(t')dt'-\dint\vMHt{2}(t')E(t')dt',	\\
\end{equation}

\label{MHt5}
\begin{equation}
\MHt{5}=
\begin{pmatrix}
-\dpsty\frac{22}{15}\Zf{1} & 0 & 0 \\
0 & \dpsty6\Zh{2} & 0 \\
0 & 0 & -\dpsty\frac{16}{3}\Zh{3} \\ 
\end{pmatrix}, 
~~~~\MHt{8}=
\begin{pmatrix}
-\dpsty\frac{1}{3}\Zf{1} & 0 & 0 \\
0 & \dpsty\Zh{2} & 0 \\
0 & 0 & -\dpsty\frac{8}{3}\Zh{3} \\ 
\end{pmatrix},
\end{equation}

\begin{figure}[t*]
\begin{center}
\includegraphics[scale=0.18]{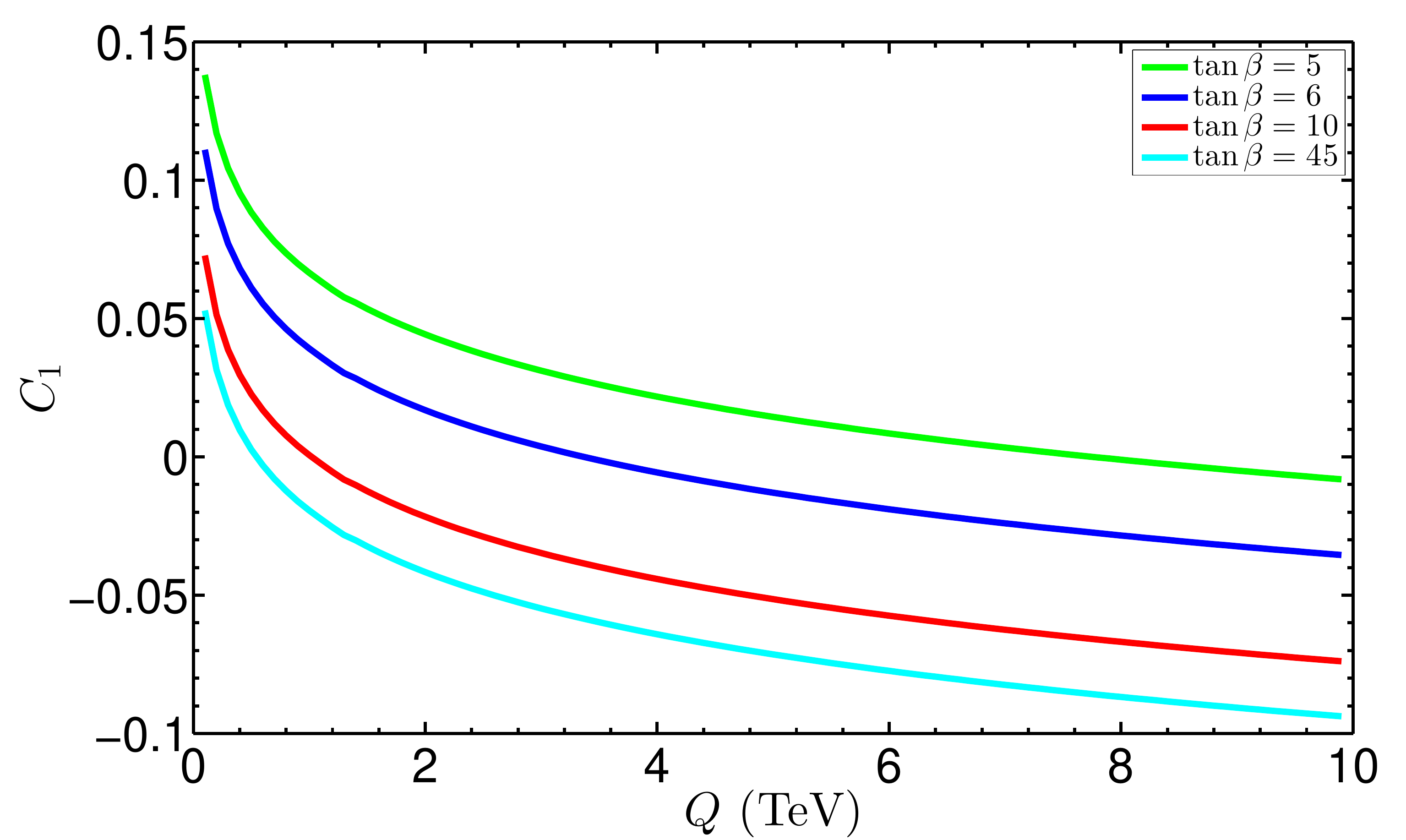}
\includegraphics[scale=0.18]{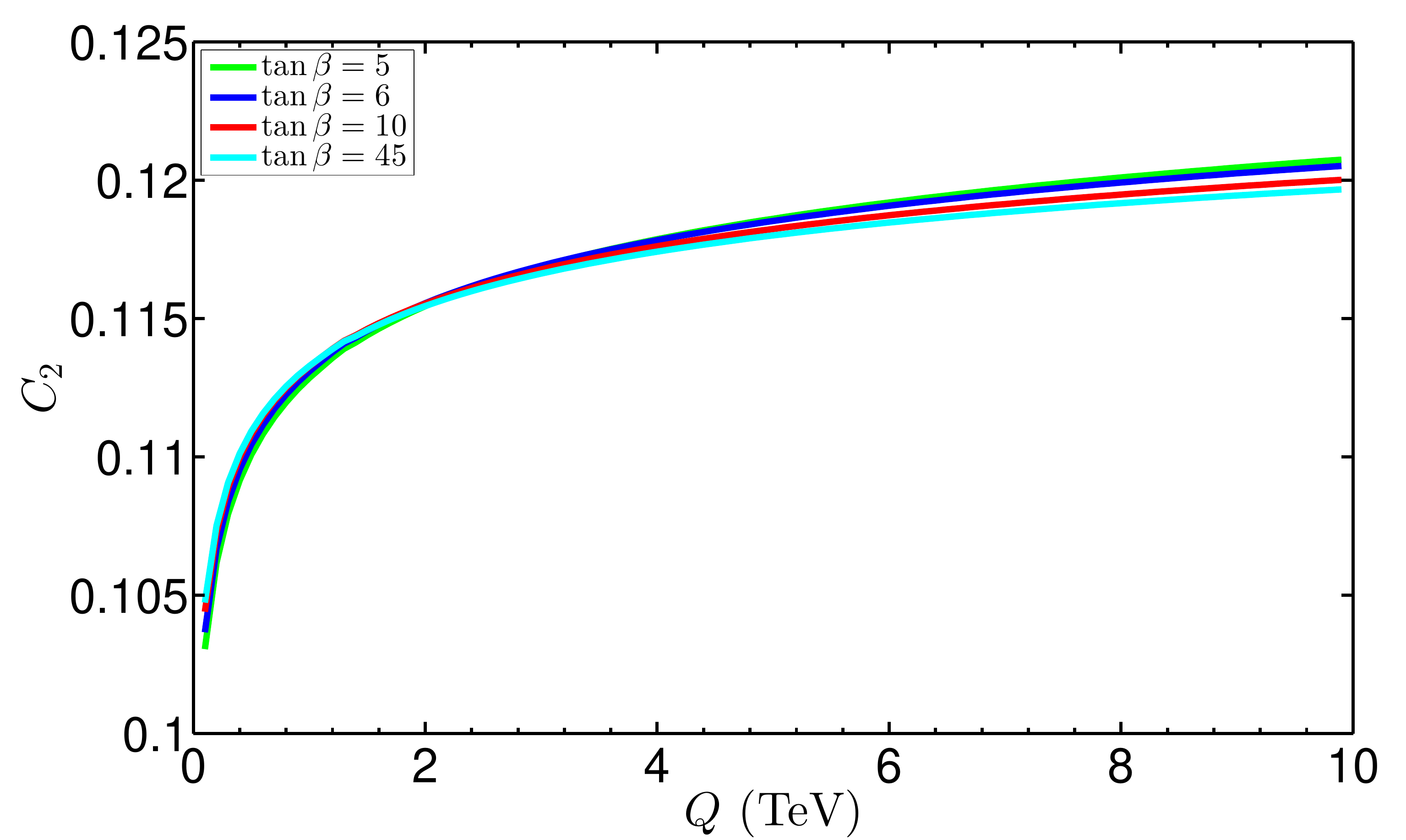}\\
\includegraphics[scale=0.18]{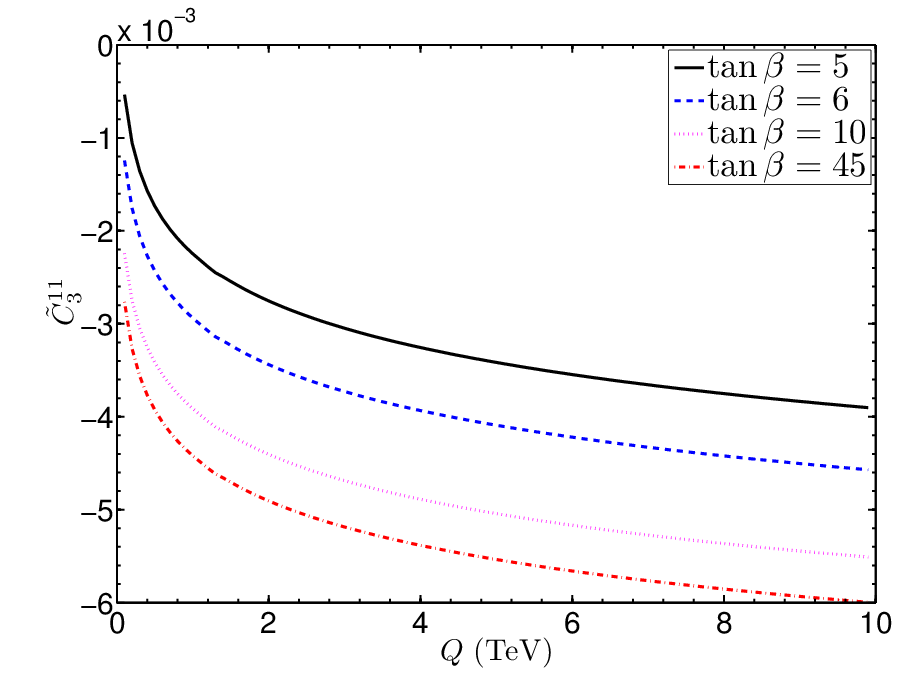}
\includegraphics[scale=0.18]{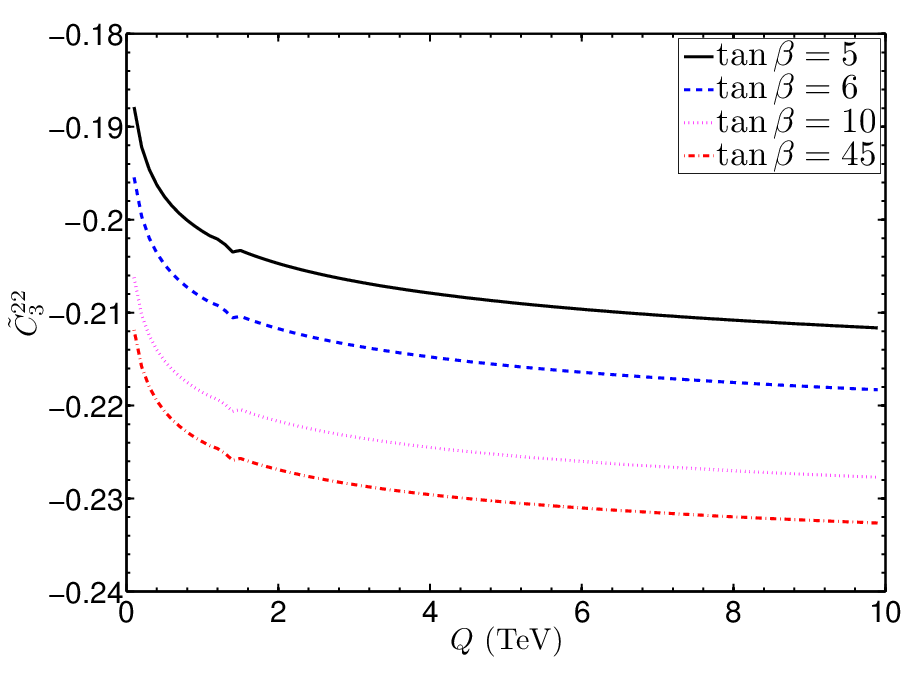}
\includegraphics[scale=0.18]{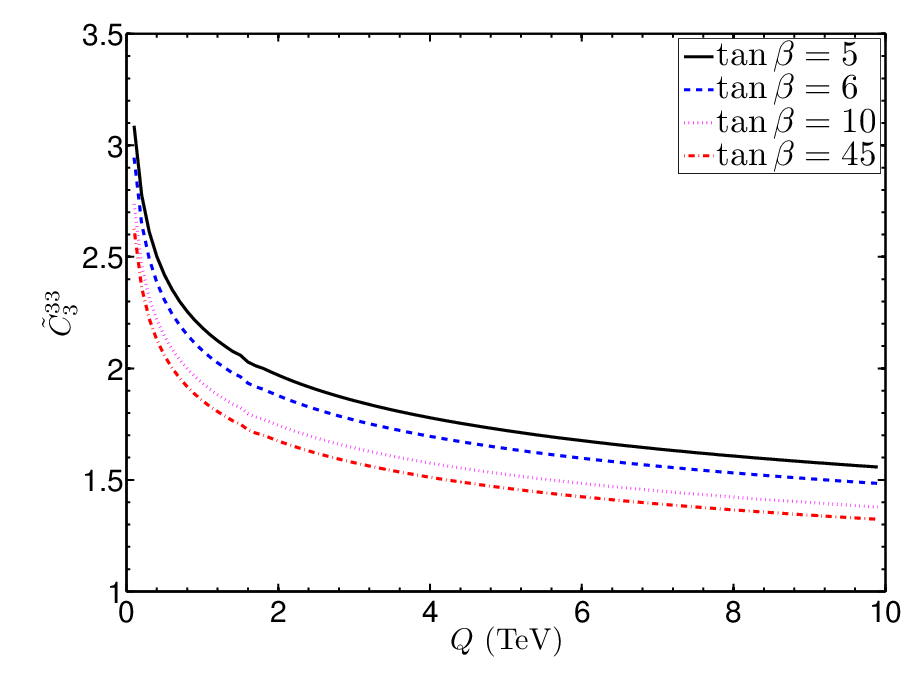}
\caption{\label{Cval}
The upper panels: RG evolution of  $C_1(Q)$ and $C_2(Q)$  as a function of the renormalization group
scale $Q$   at different $\tan\beta$. Left  panel: $C_1(Q)$ at $\tan\beta=5,6,10$ and $45$. Right  panel: $C_2(Q)$ at $\tan\beta=5,6,10$ and $45$. It is seen that $C_1(Q)$ turns negative as the scale $Q$ increases while $C_2(Q)$ remains 
positive. It is also seen that $C_1(Q)$ is very sensitive to $\tan\beta$ while $C_2$ is very insensitive to $\tan\beta$. 
The lower panels:
An exhibition of $\tilde{C}_3^{ii}$ at different $\tan\beta$. Left  panel: $\tilde{C}_3^{11}$ at $\tan\beta=5,6,10$ and $45$. Middle  panel: $\tilde{C}_3^{22}$ at $\tan\beta=5,6,10$ and $45$. Right  panel: $\tilde{C}_3^{33}$ at $\tan\beta=5,6,10$ and $45$. 
It is seen that $\tilde C_3^{11}$ and $\tilde C_3^{22}$ are negative, which allows the possibility of new focal
curves as discussed in the text. }
\end{center}
\end{figure}

\label{MHt6}
\begin{equation}
\MHt{6}=\dint\left( \vMHt{2}(t')\right)^\mathsf{T} \left( \vMHt{2}(t')\right) E(t')dt'.	\\
\end{equation}
Similarly, $\tilde{F}_i(t)$ defined by  
\begin{eqnarray}
\label{Ft2}
\Ft{2}&=&\frac{8}{15}\ft{1}+\frac{8}{3}\ft{2},	\\
\label{Ft3}
\Ft{3}&=&F(t)\Ft{2}(t)-\dint E(t')\Ft{2}(t')dt',	\\
\label{Ft4}
\Ft{4}&=&\dint E(t')\Ht{5}(t')dt',	
\end{eqnarray}
can also be written in matrix forms  so that 
$\Ft{2}\equiv\left(\MFt{2}\right)_{ij}m_im_j,  \Ft{3}\equiv\left(\MFt{3}\right)_{ij}m_im_j, 
\Ft{4}\equiv\left(\MFt{4}\right)_{ij}m_im_j$, 
with $M_{\tilde{F}_i}$ defined by 
\begin{eqnarray}
\label{MFt2}
\MFt{2}(t)=
\begin{pmatrix}
\dpsty\frac{8}{15}\Zf{1}	& 0	& 0 \\
0 &	0 & 0	\\
0 & 0 & \dpsty\frac{8}{3}\Zf{3}	\\
\end{pmatrix},	\\
\label{MFt3}
\MFt{3}(t)=F(t)\MFt{2}(t)-\dint E(t')\MFt{2}(t')dt',
\label{MFt4}
~~\MFt{4}(t)&=&\dint E(t')\MHt{5}(t')dt'	.
\end{eqnarray}

We repeat the same procedure for functions $\tilde{G}_i$ defined by 
\begin{eqnarray}
\label{Gt1}
\Gt{1} =\tilde{F}_2(t)-\frac{1}{3}\tilde{H}_2(t)^2,
\label{Gt2}
~~\Gt{2} =6\tilde{F}_3(t)-\tilde{F}_4(t)-4\tilde{H}_2(t)\tilde{H}_4(t)
+
2F(t)\tilde{H}_2(t)^2-2\tilde{H}_6(t),	
\end{eqnarray}
which could also been written as,
$\Gt{1}(t) \equiv\left(\MGt{1}\right)_{ij}m_im_j, ~~\Gt{2}(t) \equiv\left(\MGt{2}\right)_{ij}m_im_j$,
with $M_{\tilde{G}_i}$ defined by 
\begin{eqnarray}
\label{MGt1}
\MGt{1}(t)= &&\MFt{2}-\frac{1}{3}\left( \vMHt{2}\right) ^\mathsf{T}\cdot\vMHt{2},	\\
\label{MGt2}
\MGt{2}(t)= &&6\MFt{3}-\MFt{4}-4\left( \vMHt{2}\right) ^\mathsf{T}\cdot\vMHt{4}	+
2F\left( \vMHt{2}\right) ^\mathsf{T}\cdot\vMHt{2}-2\MHt{6}.
\end{eqnarray}
We return now to  $\tilde{e}(t)$ and $\tilde{f}(t)$
and write these in the  matrix form so that 
$\et(t)\equiv\left(\Met\right)_{ij}m_im_j$,  and 
$\ft(t)\equiv\left(\vMft\right)_im_i$
with
\begin{eqnarray}
\Met =  \frac{3}{2\Dt^2}\left(3\Dt\left[\MGt{1}+Y_0\MGt{2}\right] + 
\frac{1}{3}\left[\vMHt{2}+6Y_0\vMHt{4}\right]^2+\Dt^2\MHt{8}\right),
~~\vMft=-\frac{6Y_0\vMHt{3}}{D(t)^2}.
\end{eqnarray}
Using the above we can write  $\mH{2}$  in the form 
\begin{eqnarray}
\label{mH2}
\mH{2} = \left(\Met\right)_{ij}m_im_j+A_0\left(\vMft\right)_im_i	+
m_0^2h(t)-A_0^2k(t).
\end{eqnarray}
Thus using Eq.(\ref{mH1}) and Eq.(\ref{mH2})  in Eq.(\ref{musqr}), we finally have
the radiative electroweak symmetry breaking equation for non-universalities as
given in Eq.(\ref{4.1}).

\end{appendix}


\begin{thebibliography}{999}

 \bibitem{:2012gk}
 ATLAS Collaboration Collaboration (G.~Aad {\em et~al.}), {\em Phys.Lett.} {\bf
  B716}, 1  (2012), \href{http://arxiv.org/abs/1207.7214}{{\ttfamily
  arXiv:1207.7214 [hep-ex]}}.

\bibitem{:2012gu}
 CMS Collaboration Collaboration (S.~Chatrchyan {\em et~al.}), {\em Phys.Lett.}
  {\bf B716}, 30  (2012), \href{http://arxiv.org/abs/1207.7235}{{\ttfamily
  arXiv:1207.7235 [hep-ex]}}.

\bibitem{Englert:1964et}
F.~Englert and R.~Brout, {\em Phys. Rev. Lett.} {\bf 13}, 321  (1964).

\bibitem{Higgs:1964pj}
P.~W. Higgs, {\em Phys. Rev. Lett.} {\bf 13}, 508  (1964).

\bibitem{Guralnik:1964eu}
G.~Guralnik, C.~Hagen and T.~Kibble, {\em Phys. Rev. Lett.} {\bf 13}, 585
  (1964).

\bibitem{Weinberg:1967tq}
S.~Weinberg, {\em Phys.Rev.Lett.} {\bf 19}, 1264  (1967).

\bibitem{salam}
A.~Salam, Elementary paricle theory (Almqvist and Wiksells, Stockholm, 1968),
  p. 367.

\bibitem{Chamseddine:1982jx}
  A.~H.~Chamseddine, R.~L.~Arnowitt, P.~Nath,
  Phys.\ Rev.\ Lett.\  {\bf 49}, 970 (1982); 
   P.~Nath, R.~L.~Arnowitt, A.~H.~Chamseddine,
  Phys.\ Lett.\  {\bf B121}, 33 (1983).
   For  a review see   P.~Nath,
   [hep-ph/0307123].

\bibitem{Nath:1983aw}
P.~Nath, R.~L. Arnowitt and A.~H. Chamseddine, {\em Nucl. Phys.} {\bf B227},
  121  (1983).

\bibitem{Hall:1983iz}
L.~J. Hall, J.~D. Lykken and S.~Weinberg, {\em Phys.Rev.} {\bf D27}, 2359
  (1983).

\bibitem{Arnowitt:1992aq} 
  R.~L.~Arnowitt and P.~Nath,
  Phys.\ Rev.\ Lett.\  {\bf 69}, 725 (1992).

\bibitem{Akula:2011aa}
S.~Akula, B.~Altunkaynak, D.~Feldman, P.~Nath and G.~Peim, {\em Phys. Rev.}
  {\bf D85},   075001  (2012), \href{http://arxiv.org/abs/1112.3645}{{\ttfamily
  arXiv:1112.3645 [hep-ph]}}.

\bibitem{Akula:2012kk}
S.~Akula, P.~Nath and G.~Peim, {\em Phys.Lett.} {\bf B717}, 188  (2012),
  \href{http://arxiv.org/abs/1207.1839}{{\ttfamily arXiv:1207.1839 [hep-ph]}}.

\bibitem{Arbey:2012dq}
A.~Arbey, M.~Battaglia, A.~Djouadi and F.~Mahmoudi  (2012),
  \href{http://arxiv.org/abs/1207.1348}{{\ttfamily arXiv:1207.1348 [hep-ph]}}.

\bibitem{Ellis:2012aa}
J.~Ellis and K.~A. Olive, {\em Eur.Phys.J.} {\bf C72},   2005  (2012),
  \href{http://arxiv.org/abs/1202.3262}{{\ttfamily arXiv:1202.3262 [hep-ph]}}.

\bibitem{Baer:2012mv}
H.~Baer, V.~Barger, P.~Huang, D.~Mickelson, A.~Mustafayev {\em et~al.}  (2012),
  \href{http://arxiv.org/abs/1210.3019}{{\ttfamily arXiv:1210.3019 [hep-ph]}}.

\bibitem{Nath:2012nh}
P.~Nath, {\em Int.J.Mod.Phys.} {\bf A27},   1230029  (2012),
  \href{http://arxiv.org/abs/1210.0520}{{\ttfamily arXiv:1210.0520 [hep-ph]}}.

\bibitem{Carena:2002es} 
  M.~S.~Carena and H.~E.~Haber,
  Prog.\ Part.\ Nucl.\ Phys.\  {\bf 50}, 63 (2003)
  [hep-ph/0208209];
  A.~Djouadi,
  Phys.\ Rept.\  {\bf 459}, 1 (2008)
  [hep-ph/0503173].

\bibitem{Babu:2008ge}
K.~Babu, I.~Gogoladze, M.~U. Rehman and Q.~Shafi, {\em Phys.Rev.} {\bf D78},
  055017  (2008), \href{http://arxiv.org/abs/0807.3055}{{\ttfamily
  arXiv:0807.3055 [hep-ph]}}.

\bibitem{Martin:2010dc}
S.~P. Martin, {\em Phys.Rev.} {\bf D82},   055019  (2010),
  \href{http://arxiv.org/abs/1006.4186}{{\ttfamily arXiv:1006.4186 [hep-ph]}}.

\bibitem{Feng:2013mea} 
  W.~-Z.~Feng and P.~Nath,
  arXiv:1303.0289 [hep-ph].

\bibitem{Joglekar:2013zya} 
  A.~Joglekar, P.~Schwaller and C.~E.~M.~Wagner,
  arXiv:1303.2969 [hep-ph].

\bibitem{Ibanez:2007pf}
L.~Ibanez and G.~Ross, {\em Comptes Rendus Physique} {\bf 8}, 1013  (2007),
  \href{http://arxiv.org/abs/hep-ph/0702046}{{\ttfamily arXiv:hep-ph/0702046
  [HEP-PH]}}.

\bibitem{Chan:1997bi}
K.~L. Chan, U.~Chattopadhyay, and P.~Nath,
  \href{http://dx.doi.org/10.1103/PhysRevD.58.096004}{{\em Phys.Rev.}
  {\bfseries D58} (1998) 096004},
\href{http://arxiv.org/abs/hep-ph/9710473}{{\ttfamily arXiv:hep-ph/9710473
  [hep-ph]}}.

\bibitem{Chattopadhyay:2003xi} 
  U.~Chattopadhyay, A.~Corsetti and P.~Nath,
  Phys.\ Rev.\ D {\bf 68}, 035005 (2003)
  [hep-ph/0303201].

  \bibitem{bbbkt}
  H.~Baer,  C.~Balazs, A.~Belyaev, T.~Krupovnickas and X.~Tata,
  JHEP {\bf 0306}, 054 (2003).

\bibitem{Feldman:2011ud}
  D.~Feldman, G.~Kane, E.~Kuflik and R.~Lu,
  arXiv:1105.3765 [hep-ph].

\bibitem{Nath:1997qm}
  P.~Nath, R.~L.~Arnowitt,
  Phys.\ Rev.\  {\bf D56}, 2820-2832 (1997);
  J.~R.~Ellis, K.~A.~Olive, Y.~Santoso,
  Phys.\ Lett.\  {\bf B539}, 107-118 (2002).

\bibitem{Ibanez:1984vq}
  L.~E.~Ibanez, C.~Lopez and C.~Munoz,
  Nucl.\ Phys.\  B {\bf 256}, 218 (1985);
  L.~E.~Ibanez, C.~Lopez,
  Nucl.\ Phys.\  {\bf B233}, 511 (1984).

\bibitem{Arnowitt:1992qp}
  R.~L.~Arnowitt and P.~Nath,
  Phys.\ Rev.\  D {\bf 46}, 3981 (1992).

\bibitem{Feng:1999mn}
  J.~L.~Feng, K.~T.~Matchev and T.~Moroi,
  Phys.\ Rev.\ Lett.\  {\bf 84}, 2322 (2000)
  [arXiv:hep-ph/9908309].

\bibitem{Akula:2011jx} 
  S.~Akula, M.~Liu, P.~Nath and G.~Peim,
  Phys.\ Lett.\ B {\bf 709}, 192 (2012)
  [arXiv:1111.4589 [hep-ph]].

\bibitem{cmsREACH}
 [CMS Collaboration],
  arXiv:1101.1628 [hep-ex]; 
  arXiv:1109.2352 [hep-ex]; CMS-PAS-SUS-11-005; CMS-PAS-SUS-11-006; CMS-PAS-SUS-11-013; CMS-PAS-SUS-11-015.

\bibitem{AtlasSUSY}
  [ATLAS Collaboration],
  arXiv:1102.2357 [hep-ex].

\bibitem{atlas0lep}
[ATLAS Collaboration],
  arXiv:1102.5290 [hep-ex].

\bibitem{atlas165pb}
ATLAS Collaboration, 
ATLAS-CONF-2011-086.

\bibitem{atlas1fb}
  G.~Aad {\it et al.} [ ATLAS Collaboration ],
  [arXiv:1109.6572 [hep-ex]];
  arXiv:1110.2299 [hep-ex].

\bibitem{planck}
   http://www.sciops.esa.int/SA/PLANCK/docs/Planck 2013 results 16.pdf

  \bibitem{pdgrev}
  K.~Nakamura {\it et al.} [ Particle Data Group Collaboration ],
  J.\ Phys.\ G {\bf G37}, 075021 (2010).

\bibitem{bphys}
  E.~Barberio {\it et al.}
  arXiv:0808.1297 [hep-ex].

\bibitem{cmslhcbbsmumu}
  CDF~Collaboration,
  arXiv:1107.2304 [hep-ex];
 {\bf CMS} and {\bf LHCb} Collaborations.
LHCb-CONF-2011-047, CMS PAS BPH-11-019.

\bibitem{Abazov:2010fs}
  V.~M.~Abazov {\it et al.}  [D0 Collaboration],
  Phys.\ Lett.\  B {\bf 693}, 539 (2010).

\bibitem{Akula:2011ke}
  S.~Akula, D.~Feldman, P.~Nath and G.~Peim,
  arXiv:1107.3535 [hep-ph].

\bibitem{belanger}
  G.~Belanger,  et.al 
  Comput.\ Phys.\ Commun.\  {\bf 180}, 747 (2009);
  Comput.\ Phys.\ Commun.\  {\bf 182}, 842 (2011).

\bibitem{Allanach}
        B.~C.~Allanach, 
Comput.\ Phys.\ Commun.\ \textbf{143}, 305 (2002). 
Version 3.2.4 was used in this analysis.

\bibitem{Martin:2009ad}
  S.~P.~Martin,
  arXiv:0903.3568 [hep-ph].

\bibitem{Feldman:2009zc}
  D.~Feldman, Z.~Liu, P.~Nath,
  Phys.\ Rev.\  {\bf D80}, 015007 (2009).
  [arXiv:0905.1148 [hep-ph]];
  N.~Chen, D.~Feldman, Z.~Liu, P.~Nath, G.~Peim,
  Phys.\ Rev.\  {\bf D83}, 035005 (2011).
  [arXiv:1011.1246 [hep-ph]].

\bibitem{Gogoladze:2012yf} 
  I.~Gogoladze, F.~Nasir and Q.~Shafi,
  arXiv:1212.2593 [hep-ph].

\bibitem{Nath:2010zj}
  P.~Nath, B.~D.~Nelson, H.~Davoudiasl, B.~Dutta, D.~Feldman, Z.~Liu, T.~Han and P.~Langacker {\it et al.},
  Nucl.\ Phys.\ Proc.\ Suppl.\  {\bf 200-202} (2010) 185
  [arXiv:1001.2693 [hep-ph]].

\bibitem{Kaufman:2013pya} 
  B.~L.~Kaufman, B.~D.~Nelson, M.~K.~Gaillard and ,
  arXiv:1303.6575 [hep-ph].

 \bibitem{drw}
 Dimopoulos, S.Raby, and F.Wilczek, Phys. Lett. 1128, 133
(1982).

\bibitem{enr}
J. Ellis, D. V. Nanopoulos, and S.Rudaz, Nucl. Phys.
8202, 43 (1982).

\bibitem{Arnowitt:1985iy} 
  R.~L.~Arnowitt, A.~H.~Chamseddine and P.~Nath,
  Phys.\ Lett.\ B {\bf 156}, 215 (1985);
  P.~Nath, A.~H.~Chamseddine and R.~L.~Arnowitt,
  Phys.\ Rev.\ D {\bf 32}, 2348 (1985).

\bibitem{Hisano:1992jj} 
  J.~Hisano, H.~Murayama and T.~Yanagida,
  Nucl.\ Phys.\ B {\bf 402}, 46 (1993)
  [hep-ph/9207279].

\bibitem{Goto:1998qg} 
  T.~Goto and T.~Nihei,
  Phys.\ Rev.\ D {\bf 59}, 115009 (1999)
  [hep-ph/9808255].

\bibitem{Nath:2006ut} 
  P.~Nath and P.~Fileviez Perez,
  Phys.\ Rept.\  {\bf 441}, 191 (2007)
  [hep-ph/0601023].

\bibitem{Raby:2008pd} 
  S.~Raby, T.~Walker, K.~S.~Babu, H.~Baer, A.~B.~Balantekin, V.~Barger, Z.~Berezhiani and A.~de Gouvea {\it et al.},
  arXiv:0810.4551 [hep-ph].

\bibitem{Hewett:2012ns} 
  J.~L.~Hewett, H.~Weerts, R.~Brock, J.~N.~Butler, B.~C.~K.~Casey, J.~Collar, A.~de Govea and R.~Essig {\it et al.},
  arXiv:1205.2671 [hep-ex].

\bibitem{Nath:1997jc} 
  P.~Nath and R.~L.~Arnowitt,
  Phys.\ Atom.\ Nucl.\  {\bf 61}, 975 (1998)
  [Yad.\ Fiz.\  {\bf 61}, 1069 (1998)]
  [hep-ph/9708469].

\bibitem{Aoki:2006ib} 
  Y.~Aoki, C.~Dawson, J.~Noaki and A.~Soni,
  Phys.\ Rev.\ D {\bf 75}, 014507 (2007)
  [hep-lat/0607002].

\bibitem{Dermisek:2000hr} 
  R.~Dermisek, A.~Mafi, S.~Raby and ,
  Phys.\ Rev.\ D {\bf 63}, 035001 (2001)
  [hep-ph/0007213].

\bibitem{EmmanuelCosta:2003pu} 
  D.~Emmanuel-Costa, S.~Wiesenfeldt and ,
  Nucl.\ Phys.\ B {\bf 661}, 62 (2003)
  [hep-ph/0302272].

\bibitem{Nihei:1994tx} 
  T.~Nihei, J.~Arafune and ,
  Prog.\ Theor.\ Phys.\  {\bf 93}, 665 (1995)
  [hep-ph/9412325].

\bibitem{Turzynski:2001zs} 
  K.~Turzynski,
  JHEP {\bf 0210}, 044 (2002)
  [hep-ph/0110282].

\bibitem{Murayama:2001ur} 
  H.~Murayama and A.~Pierce,
  Phys.\ Rev.\ D {\bf 65}, 055009 (2002)
  [hep-ph/0108104].

\bibitem{Nath:2007eg} 
  P.~Nath, R.~M.~Syed and ,
  Phys.\ Rev.\ D {\bf 77}, 015015 (2008)
  [arXiv:0707.1332 [hep-ph]].

\bibitem{Arnowitt:1993pd} 
  R.~L.~Arnowitt and P.~Nath,
  Phys.\ Rev.\ D {\bf 49}, 1479 (1994)
  [hep-ph/9309252].

\bibitem{Babu:2010ej} 
  K.~S.~Babu, J.~C.~Pati, Z.~Tavartkiladze and ,
  JHEP {\bf 1006}, 084 (2010)
  [arXiv:1003.2625 [hep-ph]].


\bibitem{Babu:2011tw} 
  K.~S.~Babu, I.~Gogoladze, P.~Nath, R.~M.~Syed and ,
  Phys.\ Rev.\ D {\bf 85}, 075002 (2012)
  [arXiv:1112.5387 [hep-ph]];

 \bibitem{bgns}
  K.~S.~Babu, I.~Gogoladze, P.~Nath, R.~M.~Syed and ,
  Phys.\ Rev.\ D {\bf 72}, 095011 (2005)
  [hep-ph/0506312].
  K.~S.~Babu, I.~Gogoladze, P.~Nath, R.~M.~Syed and ,
  Phys.\ Rev.\ D {\bf 74}, 075004 (2006)
  [hep-ph/0607244].

\bibitem{bg} 
R. Barbieri and G. F. Giudice, Nucl. Phys. B 306, 63(1988)

\bibitem{Ciafaloni:1996zh}
P.~Ciafaloni and A.~Strumia,
  '' \href{http://dx.doi.org/10.1016/S0550-3213(97)00138-7}{{\em
  Nucl.Phys.} {\bfseries B494} (1997) 41--53},
\href{http://arxiv.org/abs/hep-ph/9611204}{{\ttfamily arXiv:hep-ph/9611204
  [hep-ph]}}.

\bibitem{Bhattacharyya:1996dw}
G.~Bhattacharyya and A.~Romanino, 
  \href{http://dx.doi.org/10.1103/PhysRevD.55.7015}{{\em Phys.Rev.} {\bfseries
  D55} (1997) 7015--7019},
\href{http://arxiv.org/abs/hep-ph/9611243}{{\ttfamily arXiv:hep-ph/9611243
  [hep-ph]}}.

\bibitem{ACR}
G.W. Anderson, D.J.Casta\~ no and A. Riotto, Phys. Rev. {\bf D55}, 2950(1997).

\bibitem{Kane:1998im}
  G.~L.~Kane and S.~F.~King,
  Phys.\ Lett.\  B {\bf 451} (1999) 113
  [arXiv:hep-ph/9810374].

\bibitem{Chankowski:1998xv}
  P.~H.~Chankowski, J.~R.~Ellis, M.~Olechowski, S.~Pokorski,
  Nucl.\ Phys.\  {\bf B544}, 39-63 (1999).
  [hep-ph/9808275].

\bibitem{Casas:2003jx}
  J.~A.~Casas, J.~R.~Espinosa, I.~Hidalgo,
  JHEP {\bf 0401}, 008 (2004).
  [hep-ph/0310137].

\bibitem{Martin:2010kk}
  S.~P.~Martin,
  Phys.\ Rev.\  D {\bf 83}, 035019 (2011).

\bibitem{ArkaniHamed:2012kq}
  N.~Arkani-Hamed, K.~Blum, R.~T.~D'Agnolo, J.~Fan and ,
  JHEP {\bf 1301} (2013) 149
  [arXiv:1207.4482 [hep-ph]].

\bibitem{Ghilencea:2012gz} 
  D.~M.~Ghilencea, H.~M.~Lee and M.~Park,
  arXiv:1203.0569 [hep-ph].

\bibitem{Jaeckel:2012aq} 
  J.~Jaeckel and V.~V.~Khoze,
  JHEP {\bf 1211}, 115 (2012)
  [arXiv:1205.7091 [hep-ph]].

\bibitem{Baer:2012cf}
H.~Baer, V.~Barger, P.~Huang, D.~Mickelson, A.~Mustafayev, {\em et al.},
\href{http://arxiv.org/abs/1212.2655}{{\ttfamily arXiv:1212.2655 [hep-ph]}}.

\bibitem{McKeen:2013dma} 
  D.~McKeen, M.~Pospelov and A.~Ritz,
  arXiv:1303.1172 [hep-ph].

\bibitem{Boehm:2013qva} 
  C.~Boehm, P.~S.~B.~Dev, A.~Mazumdar and E.~Pukartas,
  arXiv:1303.5386 [hep-ph].

\end{thebibliography}
\end{document}